# Treasure Map Toward Skyrmion Evolution in Ambient Conditions:

# A Perspective from Electronic Instabilities and the Density of Energy

*Xudong Huai and Thao T. Tran\**

AUTHOR ADDRESS

Department of Chemistry, Clemson University, Clemson, South Carolina 29634, United States

ABSTRACT

Magnetic skyrmions with topologically protected properties are anticipated to shape the future of electronics. Understanding how skyrmions may evolve in ambient conditions presents a key challenge in the pursuit of technologically significant materials. In this perspective, we focus on electronic instabilities and the density of energy of established skyrmion hosts, where a pathway to a skyrmion phase transition is readily available, to identify signposts for the emergence of skyrmions. We value the impressive research efforts in the field that have built the foundation for many more enticing breakthroughs to come. We share a framework that connects the electronic origins of skyrmion formation to the temperature and field requirements, allowing predictions of candidate materials that may host skyrmions at ambient conditions (the treasure).

## 1. Introduction

Magnetic skyrmions—nanoscale, vortex-like spin textures with topological protection—display unique physical properties, such as the topological Hall effect, magnetic-field-driven or electric-field-driven skyrmion motion, and non-reciprocal response.[1-4] These characteristics make them attractive for spintronics.[5-7] Despite remarkable research progress, realizing skyrmions under practical conditions, stable at room temperature and operable in low magnetic or electric fields, remains a significant challenge. While some systems can host skyrmions, they often require cryogenic conditions or high external fields, hindering their potential value in shaping the future of electronics.

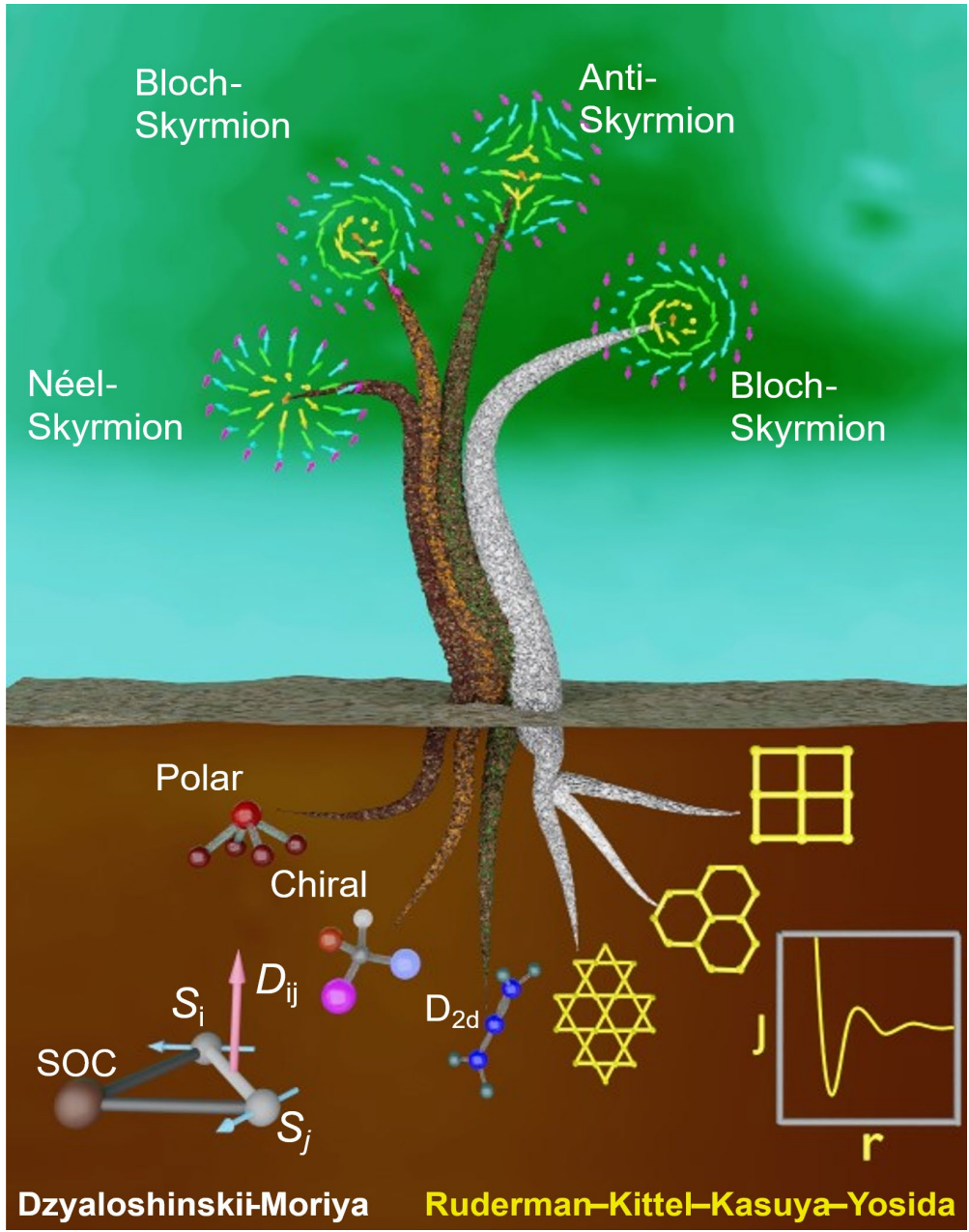

**Figure 1.** The Skyrmion tree showing the formation mechanisms. The roots of the tree indicate the underlying driving force. Dzyaloshinkii-Moriya interactions, facilitated by larger spin-orbital coupling, stabilize skyrmions in non-centrosymmetric systems. Ruderman-Kittel-Kasuya-Yosdia exchange interactions, assisted by geometrically frustrated lattice systems, stabilize skyrmions in centrosymmetric metallic systems.

**Underlying magnetic exchange interactions**

The emergence of skyrmions can be understood from two fundamental mechanisms, each rooted in distinct electronic and structural conditions (**Figure 1**). On one side, skyrmions evolve from the Dzyaloshinskii–Moriya (DM) interaction, which can be fostered in noncentrosymmetric (NCS) lattices with strong spin–orbit coupling.[8-13] Depending on the local site symmetry, different types of skyrmions can emerge: polar point-group symmetries favor Néel-type skyrmions; chiral symmetries stabilize Bloch-type skyrmions; and $D_{2d}$ symmetry enables the formation of anti-skyrmions. These three symmetry-derived roots lead to distinct spin textures that collectively populate the "DM branch" of the skyrmion family tree. On the other side, the Ruderman–Kittel–Kasuya–Yosida (RKKY) mechanism operates in centrosymmetric metallic systems, where conduction electrons mediate oscillatory exchange interactions.[14-16] In geometrically frustrated magnetic lattices—such as square, honeycomb, or kagome networks— this oscillatory coupling bends spin orientations across the lattice, producing Bloch-type skyrmions without requiring inversion symmetry breaking. These frustrated-lattice roots define the "RKKY branch," providing a parallel path to topological spin textures through electronic frustration rather than structural asymmetry.

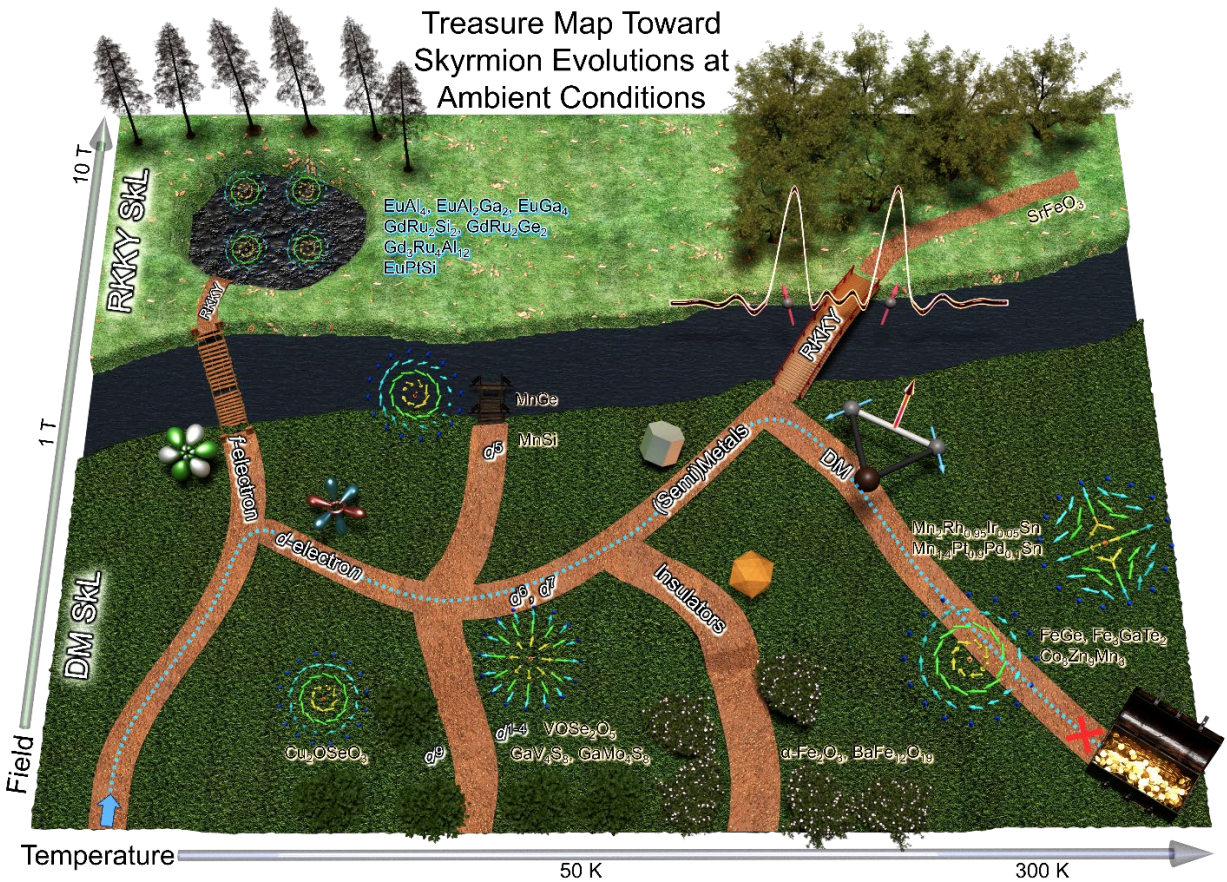

**Figure 2.** A treasure map illustrating pathways toward the realization of skyrmions at low-field and high-temperature towards ambient conditions, guided by key design parameters: *d*- or *f*-electrons, orbital occupancy, metallic or insulating character, and DM or RKKY interactions. The highlighted route marks the parameter combination leading to the target ambient skyrmion (treasure).

### Current stage of knowledge

The DM and RKKY interactions lay a mechanistic foundation upon which a treasure map for hunting materials hosting skyrmions at low-field and high-temperature towards ambient conditions can be built (**Figure 2**). The layout of the treasure map is drawn from a combination of (i) the current stage of knowledge of skyrmions enabled by elegant work in the field,[17-39] and (ii) insights gained from our electronic structure and density of energy analysis. In this landscape, the realization of skyrmions at low-field and high-temperature toward ambient conditions can be conceptualized as an adventure through a multidimensional materials design space (**Figure 2**). Let's get started. When you get to the first fork in the road, you can choose to take either the $d$- or $f$-electron pathway. The $f$-electron path takes you to a set of skyrmion hosts, such as $EuGa_4$, $GdRu_2Si_2$, and $EuPtSi$. These $4f$ magnetic metals, possessing large magnetic moments, geometrically frustrated lattices, and large exchange stiffness, display small (~a few nm) Bloch-type RKKY-driven skyrmions at low temperatures and high magnetic fields (> 1 T).[40-43] One ought to cross the "1 Tesla river" on the far left in the treasure map (low-temperature region) to arrive at this set of $4f$ skyrmion hosts. By contrast, the $d$-electron path takes you to more skyrmion host options. This rich behavior can be attributed to different $d$-electron fillings, appreciable orbital overlap, and DM exchanges. As a result, skyrmion materials with $d$ electrons typically feature large skyrmion sizes (hundreds of nm) at lower fields (< 1T) and higher temperatures.[44-51] Thus, one can enjoy exploring the DM-driven skyrmion space without crossing the 1 Tesla river. When you take the $d$-electron path and get to the next fork in the road, you can decide your next step based on $d$-electron fillings and orbital occupancy, which fundamentally shape the band structure, magnetic anisotropy, and competing magnetic exchange interactions of the DM-enabled skyrmion materials. For materials with half-filled $d^5$ shells, e.g., MnSi and MnGe, skyrmions in the 3–20 nm range evolve at 0.12 and 2.4 T, 30 and 170 K, respectively. $Cu_2OSeO_3$ ($d^9$) stabilizes Bloch-type skyrmions (~60 nm) at 0.035 T and 59 K. Materials with less than half-filled ($d^{1-4}$) configurations, such as $VOSe_2O_5$, $GaV_4S_8$, $GaMo_4S_8$ yield large Néel-type skyrmions (~140 nm) at 0.003 T $\leq \mu_0 H \leq$ 0.1 T and 7 K $\leq T \leq$ 18 K—lower fields than those of other aforementioned skyrmion materials but still at low temperatures. However, the temperature condition of skyrmion evolutions is approaching room temperature as one follows the path towards more-than-half-filled shells ($d^6$, $d^7$).[47, 52-55] A further distinction lies between metallic and insulating skyrmion hosts. Insulators, such as $\alpha$-$Fe_2O_3$, $BaFe_{12}O_{19}$,[56-58] can stabilize skyrmions at relatively low fields and closer to room temperature (0.05 T $\leq \mu_0 H \leq$ 0.2 T and 240 K $\leq T \leq$ 270 K), while metals and semimetals typically can display skyrmions at higher temperatures. Examples include FeGe (Bloch-type, ~70 nm), $Fe_3GaTe_2$ (Bloch-type, ~100 nm), and $Mn_2Rh_{0.95}Ir_{0.05}Sn$ and $Mn_{1.4}Pt_{0.9}Pd_{0.1}Sn$ (anti-skyrmions >130 nm) host the topological nontrivial spin textures close to ambient conditions (0.015 T $\leq \mu_0 H \leq$ 0.165 T and 276 K $\leq T \leq$ 400 K).[59-63] Crossing the 1 Testa river on the far right in the treasure map takes one back to the RKKY land, but with the unique, complex 4**q** helical skyrmions host—$SrFeO_3$ at a high field (10 T).[37-39, 50] Taken together, all the realizations point to some connections between the DM/RKKY exchange and the topology, size, and formation condition of skyrmions.

The skyrmion landscape built upon prior elegant studies in the field enables us to have a framework in mind as to how far we have come. But where are we going from here? How can we get to the treasure—materials hosting skyrmions at ambient conditions? Is there any other hint or insight into electronic structures that we can connect to the critical field and temperature conditions at which skyrmions evolve? We attempt to answer these questions by looking into the electronic origins that give rise to the phase transition to skyrmions.[64-69] Drawing from the current set of known skyrmion hosts, we developed a unified framework that emphasizes the role of electronic instabilities, particularly features in density of energy (DOE) and spin density of energy ($DOE_s$), as key descriptors of the conditions under which skyrmions emerge. Within this framework, we find that there is a common, distinctive feature in the $DOE_s$: a destabilizing peak at the Fermi level followed by a local minimum just above it. When normalized by unit-cell volume, these energy descriptors reveal some correlation with the critical field and temperature conditions for skyrmion evolution. This normalization can be justified since it captures the crystal structure and bonding of the skyrmion materials.

By tying the DOE and $DOE_s$ to the experimental conditions of skyrmion formation, this framework provides a powerful tool for predicting new skyrmion materials and designing skyrmions under ambient conditions. We validate this semi-quantitative model and discuss its application in collaboration with electron fillings, magnetic exchange interactions, and bonding character. We then propose how these design parameters can be integrated to realize materials hosting skyrmions towards ambient conditions in the treasure map.

## 2. Conceptual Framework and Semiquantitative Model

### 2.1. Formulation of the Density of Energy and Spin Density of Energy Approach

One emerging theme in recent studies is the role of electronic instabilities as a potential driving force for skyrmion formation.[70, 71] In particular, Fermi surface nesting, where large, parallel regions of the Fermi surface enhance susceptibility at specific wavevectors, has been identified as a key indicator of instability in several metallic skyrmion hosts. Examples include metallic systems such as $EuAl_4$, $GdRu_2Si_2$, and $Gd_2PdSi_3$.[72-75] In the $GdRu_2X_2$ (X = Si, Ge) family, an increasing degree of nesting correlates with skyrmion formation at progressively lower fields and higher temperatures. This suggests that electronic instabilities promote a phase transition to skyrmions. However, many known insulating skyrmion hosts are left out by this approach, since they lack a well-defined Fermi surface or nesting conditions. This raises the need for a broader descriptor that captures electronic instabilities across both metallic and insulating systems.

A natural extension is to consider chemical bonding, particularly the population of antibonding states near or at the Fermi level ($E_F$). If antibonding dominates near $E_F$, the system may be structurally and electronically unstable, thereby predisposing it to symmetry-breaking transitions such as skyrmion formation. This can be probed using the crystal orbital Hamilton population (COHP) method.[76-78] In this formalism, a negative sign in the projected COHP (-pCOHP) indicates antibonding (destabilizing) interactions—destructive interference of the overlap of the interacting atomic wavefunctions.[79,80] With spin polarization (non-restricted spin), spontaneous magnetization causes the spin-up (majority-spin) electrons to go down in energy and the spin-down (minority-spin) electrons to go up in energy.[81, 82, 83, 84, 85] This electronic symmetry-breaking is simply a result of bonding stabilization, alleviating electronic instabilities and stabilizing the system. While COHP is highly effective for identifying local, pairwise bonding interactions, it does not account for on-site atomic energy contributions. To complement COHP, we employed the concept of density of energy (DOE), which captures both interatomic and on-site atomic energy contributions as it integrates the entire electronic band structure with respect to energy.

The DOE is calculated separately for each spin channel (Table S1). The spin density of energy $DOE_s = DOE_{Dn} - DOE_{Up}$ reveals the difference in energies between the two spin sublattices—the exchange splitting. The corresponding $DOE_s$ integral ($E_{spin}$, Equation 1) can provide a semiquantitative guide for the link between the electronic origins and the critical field and temperature of skyrmions formation.

$$E_{spin} = \int\limits_{-\infty}^{E_F} \left(DOE_{Dn} - DOE_{Up}\right) dE \qquad (1)$$

This measure captures the net imbalance in the spin-up and spin-down channels, providing a key descriptor of electronic instabilities that can give rise to a magnetic phase transition. Within this framework, core electrons, structures, and other non-spin components contribute equally to both spin channels and cancel out upon subtraction, leaving only magnetism-specific contributions in $E_{spin}$. In addition, the set of material examples used in this analysis is skyrmion hosts; it means a pathway through which a magnetic phase transition to skyrmions is readily available. Since the magnetic contribution is primarily governed by valence electrons near $E_F$, the energy window for integration is carefully selected based on the valence orbital character of the magnetic atom (e.g., $d$- or $f$-orbitals). This can be validated from the DOE$_s$ plot, where the DOE$_s$ should be localized to the energy range of the valence orbitals. A near-zero DOE$_s$ at lower energies confirms that non-magnetic states are properly excluded from the magnetic contribution.

## 2.2. Case Study I: GdRu₂X₂ (X = Si, Ge)

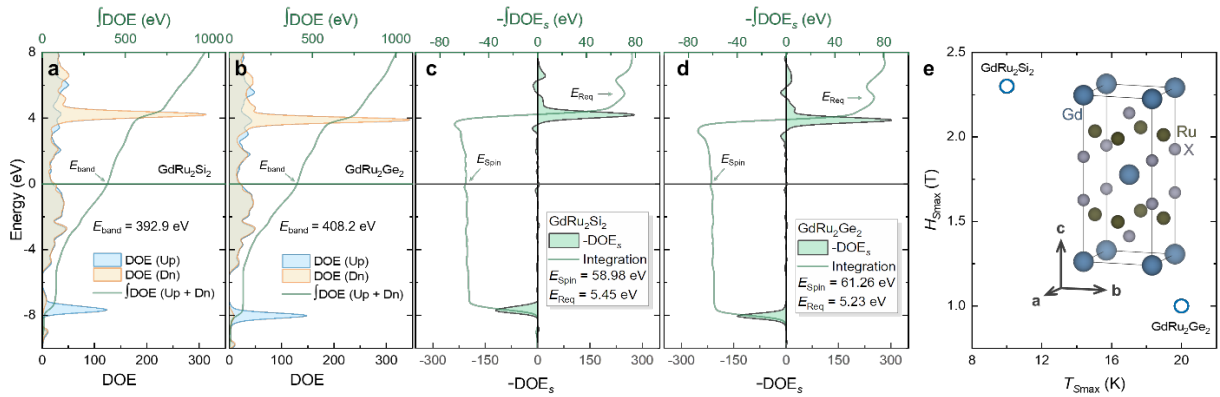

**Figure 3.** Density of energy analysis of (a) GdRu$_2$Si$_2$, (b) GdRu$_2$Ge$_2$; (c, d) Corresponding spin density of energy DOE$_s$; (e) Field and temperature conditions for skyrmion formation (Inset showing the crystal structure)

In the case of GdRu$_2$X$_2$ (X = Si and Ge), the integrated DOE ($E_{band}$) is positive at $E_F$, indicating a destabilizing contribution and an electronic instability (**Figure 3**, S1, S2).[36, 86-88] This instability can give rise to a magnetic phase transition to skyrmions. Since skyrmions emerge under finite temperature and field, the system must transition into an excited state. Thus, electronic states above $E_F$ should give us some hint of skyrmion formation. It is tricky to isolate the magnetic contribution in the DOE curves (**Figure 3a-b**) since there are strongly destabilizing energy contributions (positive DOE values and $E_{band}$). However, the DOE$_s$ curves provide better clarity. The integrated DOE$_s$ reveal changes in slope and sign at ~4 eV, consistent with the anomaly in the DOE at the same energy. Note that the change in sign of the integrated DOE$_s$ in **Figure 3c-d** is only due to the mathematical operation (Equation 1). The spin energy ($E_{spin}$, Equation 1) can give us a hint of a spin instability that can induce a phase transition to skyrmions. The integrated DOE$_s$ also reveals another unique feature—a local minimum right above 4 eV where its first derivative is zero. This local minimum is indicative of the energy required ($E_{req}$) to stabilize the system in an excited state. The finite field and temperature necessary for skyrmion formation may be linked to $E_{req}$. The system may exhibit its most pronounced magneto-entropy and topological Hall effect, which are hallmarks of the skyrmion phase, at the characteristic temperature $T_{Smax}$ and magnetic field $H_{Smax}$—conditions related to $E_{req}$. GdRu$_2$Si$_2$ hosts skyrmions at $T_f$ = 38 K, $H_f$ = 0.5 T, and its highest entropy region is at $T_{Smax}$ = 25 K, $H_{Smax}$ = 2 T, while the Ge material displays skyrmions at $T_f$ = 25 K, $H_f$ = 1.9 T, and its highest entropy region is at $T_{Smax}$ = 20 K, $H_{Smax}$ = 1.2 T (lower temperature and field compared to the Si version.[35] This can be linked to GdRu$_2$Ge$_2$ possessing a lower $E_{req}$ (5.23 eV) than GdRu$_2$Si$_2$ (5.45 eV).

## 2.3. Case Study II: EuAl$_4$, EuAl$_2$Ga$_2$, and EuGa$_4$

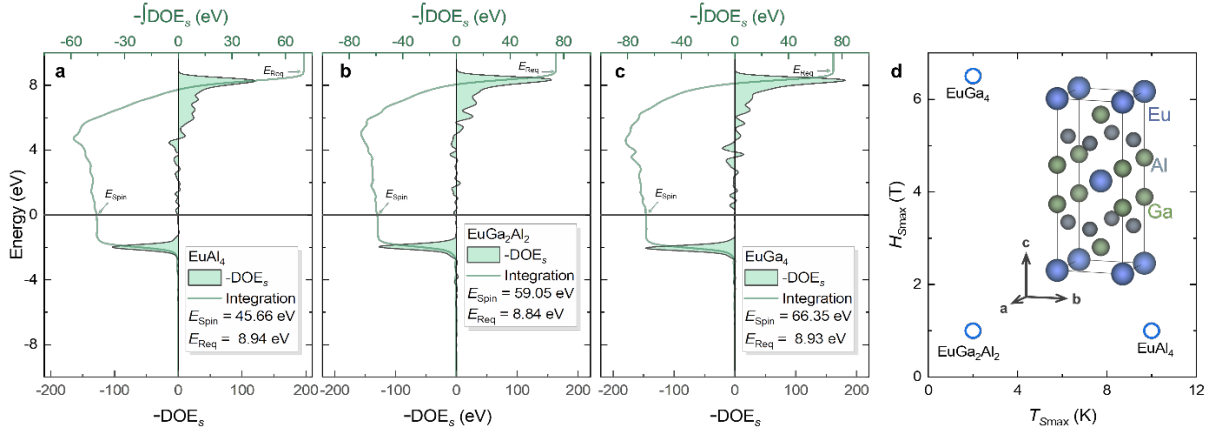

**Figure 4**. Spin density of energy DOE$_s$ of (a) EuAl$_4$, (b) EuAl$_2$Ga$_2$, (c) EuGa$_4$; (d) Field and temperature conditions for skyrmion formation (Inset showing the crystal structure)

Now, let us have a look at the series of EuAl$_4$, EuAl$_2$Ga$_2$, and EuGa$_4$—isostructural centrosymmetric skyrmion hosts (**Figure 4**, S3-5). Both EuAl$_4$ and EuAl$_2$Ga$_2$ display their largest magnetic entropy region at a similar field $H_{Smax} \approx 1$ T, but at different temperatures $T_{Smax}$.[28, 29, 31] Since EuAl$_4$ possesses a higher $E_{req}$ (8.94 eV) than EuAl$_2$Ga$_2$ (8.84 eV), more energy is required to transition the system to the excited state; thus, EuAl$_4$ has a higher $T_{Smax}$. Similarly, EuAl$_2$Ga$_2$ and EuGa$_4$ share a similar $T_{Smax} \approx 2$ K, but require different fields.[30] EuGa$_4$, which has a higher $E_{req}$ (8.93 eV), requires a larger $H_{Smax}$ (6.5 T) for its skyrmion formation.

## 2.4. Case Study III: MnSi, MnGe and FeGe

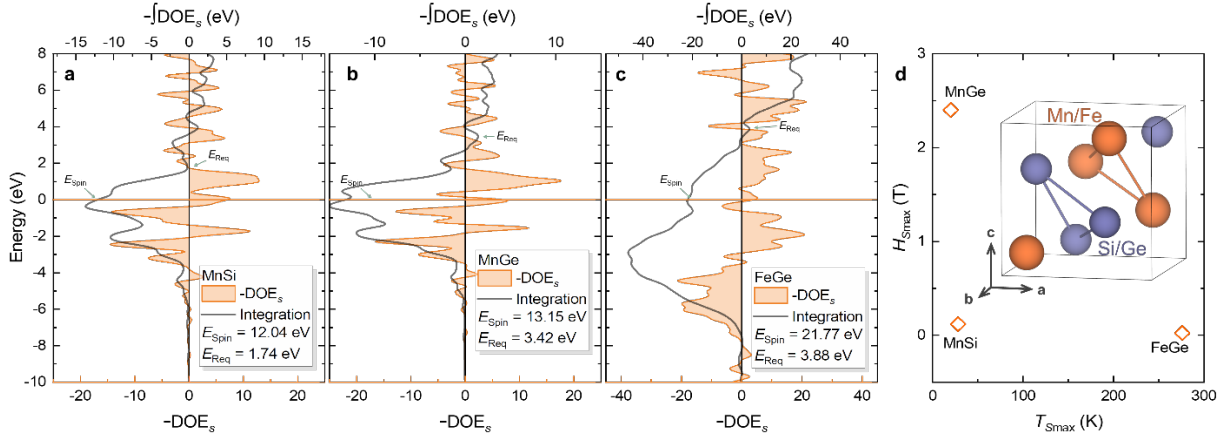

**Figure 5**. Spin density of energy DOE$_s$ of (a) MnSi, (b) MnGe, (c) FeGe; and (d) Field and temperature conditions for skyrmion formation (Inset showing the crystal structure)

In the first two case studies, we demonstrate that the semiquantitative model is effective in linking the spin density of energy to the skyrmion formation conditions in RKKY-driven systems. To further examine this model, we study skyrmion hosts facilitated by DM interactions. Let us focus on MnSi, MnGe, and FeGe, which crystallize in the noncentrosymmetric B20 crystal structure (**Figure 5**, S6-8) .[23, 24, 89-91] In these DM-driven skyrmion materials, the DOE$_s$ and corresponding integrated DOE$_s$ curves are more diffused, consistent with the more directional and spatially diffused nature of $d$-orbitals, as opposed to the more localized feature of the $f$-electron RKKY systems discussed earlier. Despite these differences, the isostructural DM-driven skyrmion hosts show similar trends: a positive $E_{spin}$ indicates a spin instability around $E_F$; and a local minimum appears above $E_F$, indicative of $E_{req}$ required to stabilize the system in an excited state. Among the selected candidates, MnSi has the lowest $E_{req}$ = 1.74 eV, followed by MnGe ($E_{req}$ = 3.42 eV) and FeGe ($E_{req}$ = 3.88 eV). This agrees well with experimental observations that MnGe requires a higher magnetic field than MnSi, and FeGe requires a higher temperature than MnSi, to enter the skyrmion phase.

## 2.5. Case Study IV: VOSe₂O₅, GaMo₄S₈, and Cu₂OSeO₃

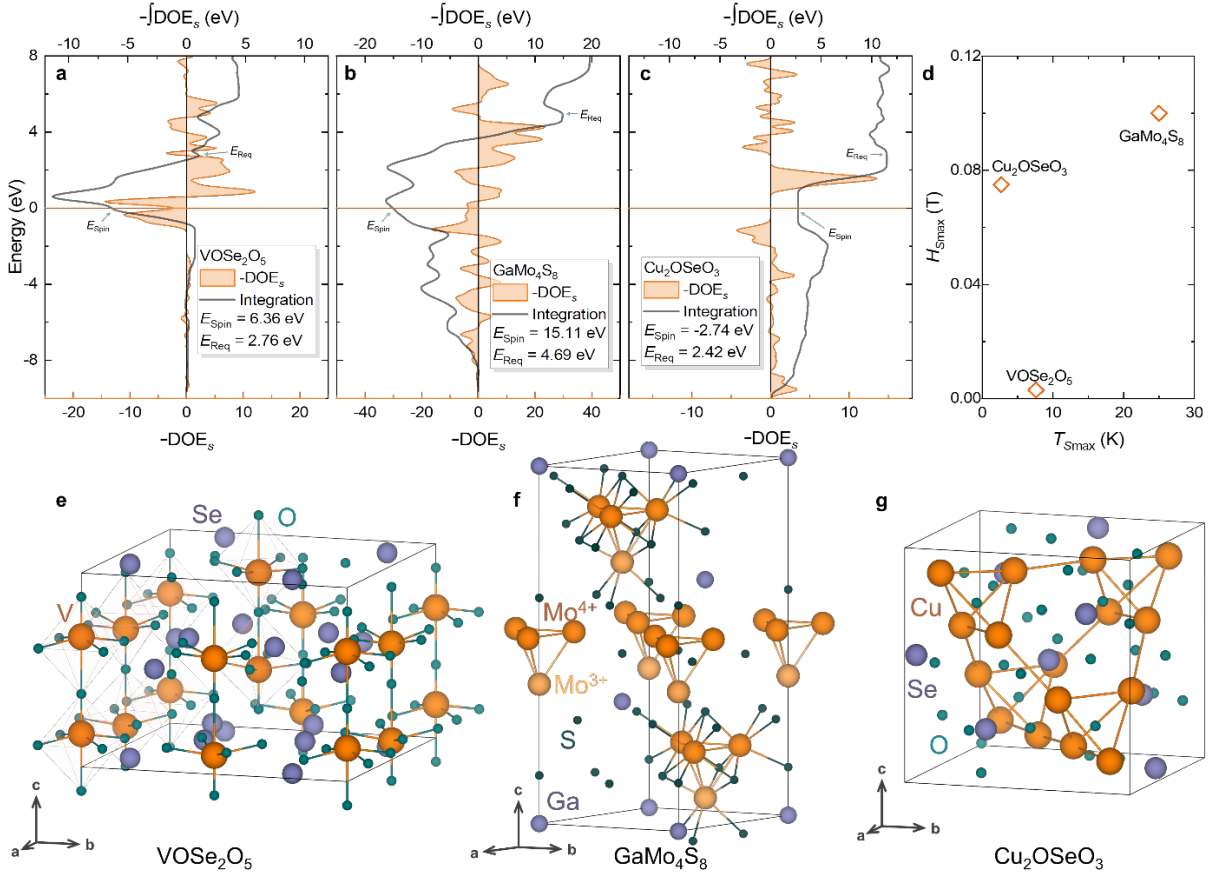

**Figure 6**. Spin density of energy DOE$_s$ of (a) VOSe₂O₅, (b) GaMo₄S₈, (c) Cu₂OSeO₃; (d) Field and temperature conditions for skyrmion formation; (e-g) Crystal structures of the skyrmion hosts.

DM-driven skyrmion materials are not limited to metallic or semimetal systems; skyrmions can also emerge in insulating compounds when the magnetic sublattice adopts a polar or chiral symmetry. Examples include VOSe₂O₅, GaMo₄S₈, and Cu₂OSeO₃ (**Figure 6**, S9-11).[17-19, 26, 27] The DOE$_s$ and corresponding integrated DOE$_s$ curves for these materials show similar trends to those of the B20 compounds. One exception is Cu₂OSeO₃, which does not exhibit a positive $E_{spin}$ like the other skyrmion hosts. This can be attributed to the electronic configuration of $Cu^{2+}$ ($3d^9$), where most 3d orbitals are filled and thus the $d$-orbital states are deep down in energy -12 eV.

(Figure S11) Nevertheless, changes in slope in the energy region between -2 eV and 3 eV are apparent, similar to the trend of the other skyrmion materials. However, it can be tricky to correlate the $E_{req}$ values of $VOSe_2O_5$, $GaMo_4S_8$, and $Cu_2OSeO_3$ and their skyrmion formation conditions (**Figure 6d**, S12-13). This challenge is due to these materials possessing different crystal structures. In order to make a more meaningful comparison, we attempt to figure out plausible ways to normalize $E_{req}$ with respect to (i) magnetic atoms or (ii) a unit cell volume. Simply normalizing by the number of magnetic atoms is insufficient because it overemphasizes the magnetic ions' contribution to $E_{req}$, neglecting the role of nonmagnetic atoms that also participate in exchange pathways and influence collective electronic and magnetic contributions. A more justified physical approach is to normalize $E_{req}$ by unit cell volume, since the relevant electronic and magnetic contributions scale with the spatial extent of the lattice. With this normalization, the $E_{req}$ values become 0.022 eV/$Å^3$ for $VOSe_2O_5$, 0.054 eV/$Å^3$ for $GaMo_4S_8$, and 0.082 eV/$Å^3$ for $Cu_2OSeO_3$. This trend explains the experimental observation that $VOSe_2O_5$ requires the lowest field and temperature for skyrmion formation among the series.

## 2.6. Generalization Across the Diverse Set of Known Skyrmion Materials

We then apply the normalization of $E_{spin}$ and $E_{req}$ with respect to a unit cell volume to a larger set of skyrmion hosts (**Figure 7**). The selected materials span a diverse range, including all DM- and RKKY-driven skyrmion hosts from metallic to insulating magnets with $d$- and $f$-electrons and various crystal structures. As shown in **Figure 7a**, although $E_{spin}$/$Å^3$ points to some hints of the evolution conditions of skyrmions, it is not sufficient to capture the trend for all the skyrmion hosts of investigation (Table S2). For example, the $E_{spin}$/$Å^3$ values of MnSi and MnGe are comparable, but the skyrmion phase in MnGe emerges at a higher field.

Now, let us turn to $E_{req}/Å^3$, which enables a more effective link to the skyrmion formation conditions. **Figure 7b** clearly reveals that systems with larger $E_{req}/Å^3$ require higher temperatures and/or magnetic fields to transition to a skyrmion state.

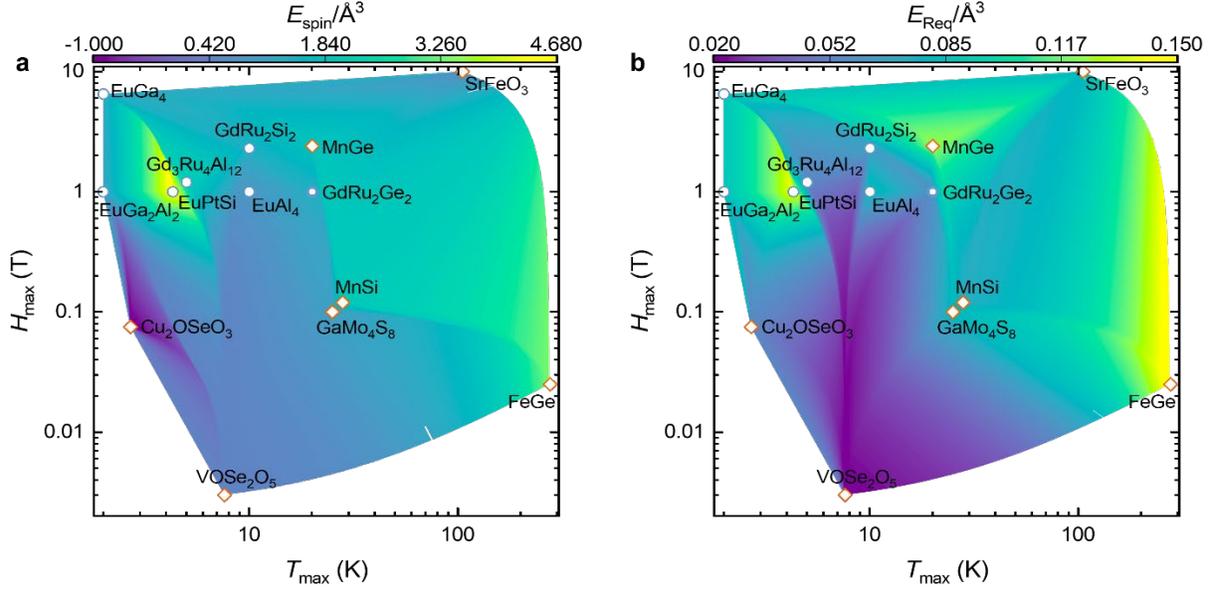

**Figure 7**. Three-dimensional plot showing the correlation between characteristic temperature ($T_{Smax}$), magnetic field ($H_{Smax}$), and the extracted (a) $E_{spin}/Å^3$ and (b) $E_{req}/Å^3$. Orange diamonds indicate $d$-electron skyrmion materials, while blue circles represent $f$-electron skyrmion hosts.

From all the case studies and the overall trend of the known skyrmion hosts, we demonstrate the key role of electronic instabilities and density of energy in manifesting the temperature and field conditions of skyrmion evolution. By examining $E_{req}/Å^3$, we can directly compare different skyrmion materials and establish a trend between their electronic origins and their skyrmion formation conditions. While this realization is enticing, it is worthwhile assessing the semi-quantitative model to see whether any potential pitfalls or improvements should be taken into account.

## 3. Assessment of the Semiquantitative Model

To evaluate the $E_{req}/Å^3$ relationship with skyrmion formation conditions, we developed a simple theoretical model that combines phonons from the Debye model and magnetic energy contribution from applied fields (Equation S1). This approach captures the proper energy scale for low-temperature, low-field skyrmion hosts, but tends to underestimate phonon contributions and overestimate field effects for high-temperature and high-field systems. (Figure S19)

To improve approximation, we constructed a second model based on experimental observables, including magneto-entropy and magnetization (Equation S2). This data-driven formulation reproduces the required energy window ($10^{-2}$–$10^{-1}$ eV/$Å^3$) more reliably and points to a convergence between this second model and the applied, semiquantitative one (Figure S20).

Although the convergence lends some credibility to the semiquantitative model of linking $E_{req}/Å^3$ with skyrmion formation conditions, some small differences remain due to the limitation on collinear ground-state electronic structure calculations. One wants to keep in mind that most skyrmion hosts exhibit noncollinear or helical spin structures at the ground state. More rigorous noncollinear supercell calculations could yield higher accuracy, though at high computational cost. Nevertheless, the presented model provides a practical framework for linking electronic instabilities and density of energy to the temperature and field conditions of skyrmion evolution and for guiding rapid materials screening using electronic structure calculations.

## 4. Materials Design Perspectives and Guiding Principles

Within this framework, we now look back at the $DOE_s$ and corresponding $DOE_s$ integral plots to gain some further insights. In addition to the case studies mentioned above, we add a few more model systems (**Figure 8**) that help build a more complete picture for our center question: how to design a material that hosts skyrmions close to or at ambient conditions—room temperature and low-to-zero field.

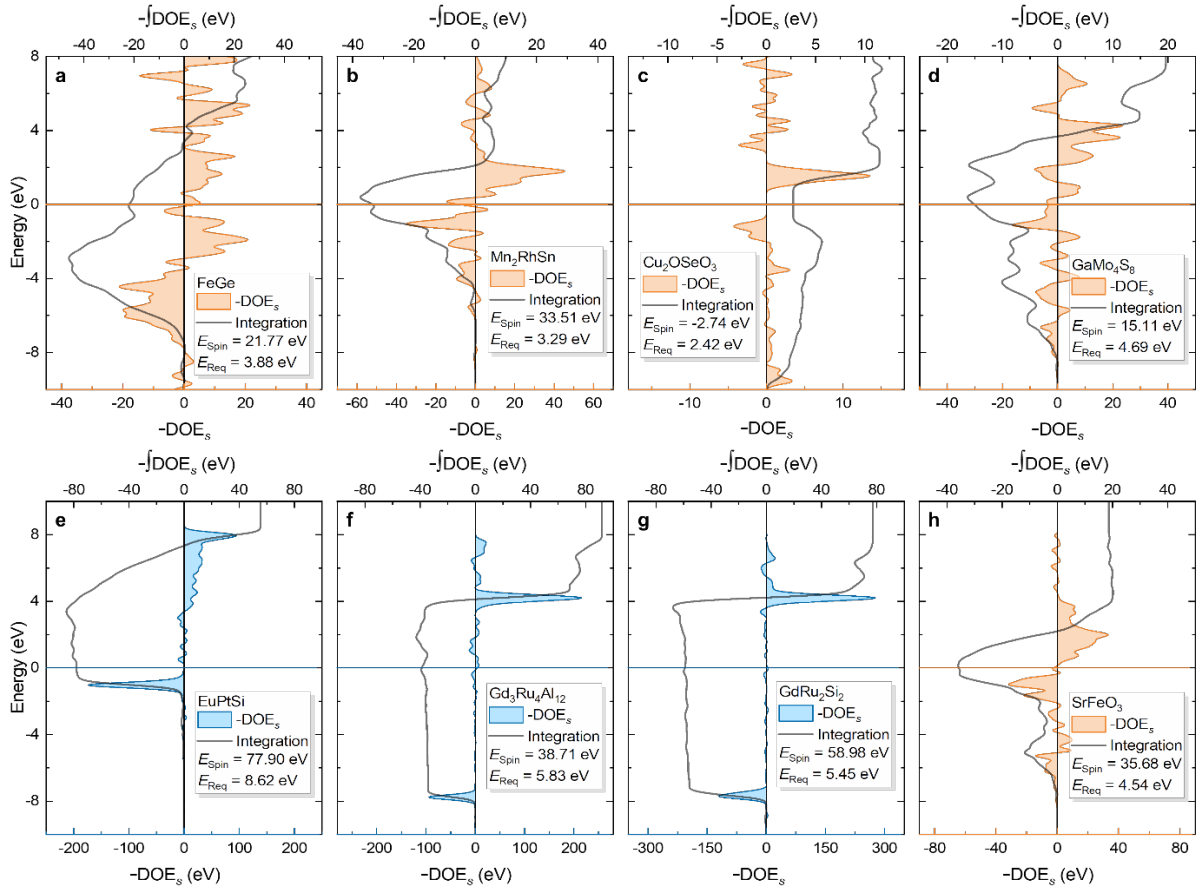

**Figure 8**. Spin density of energy DOE$_s$ of (a) FeGe, (b) Mn$_2$RhSn, (c) Cu$_2$OSeO$_3$, (d) GaMo$_4$S$_8$, (e) EuPtSi, (f) Gd$_3$Ru$_4$Al$_{12}$, (g) GdRu$_2$Si$_2$, and (h) SeFeO$_3$. Orange color denotes $d$-electron skyrmion hosts, while blue color signifies $f$-electron skyrmion materials.

## 4.1. Distinctive Behaviors of $d$- and $f$-Electron Systems

As illustrated in **Figure 8**, $d$-electron and $f$-electron skyrmion materials exhibit significantly different features in DOE$_s$. For $d$-electron systems, the DOE$_s$ curves are more dispersed, consistent with the more effective overlap and diffused character of $d$-orbitals. This trend is also reflected in **Figure 7**, where *d*-electron compounds tend to stabilize skyrmions under relatively low magnetic fields and across a wide temperature range. In contrast, *f*-electron systems generally require higher fields and lower temperatures to access their skyrmion phase.[33, 34] This difference arises partly from how magnetic moments respond to temperature and magnetic field perturbations. From magnetization measurements, where the moment *M* is a function of both temperature and field, $M(T, H)$, it is often observed that the magnetic moment varies significantly with field in $M(H)$ curves, while remaining nearly constant across temperature in $M(T)$ data. This suggests that magnetic fields more directly affect spin orientation through dipole interactions, while temperature influences spin behavior more indirectly, such as through entropy-driven fluctuations. In the context of skyrmions, this distinction is crucial: temperature supports skyrmion formation by providing entropic stabilization, while a magnetic field drives spin arrangement. As a result, skyrmions in *d*-electron magnets are more likely to emerge under more accessible (low-field and high-temperature) or ambient conditions.

### 4.2. Electron Fillings of Valence Orbitals

In addition to valence electron characters, electron fillings of valence orbitals play a key role in determining whether a system can facilitate skyrmion formation. For instance, when the valence orbitals are less than half-filled—as in $VOSe_2O_5$ and $GaMo_4S_8$—the *d*-states lie mostly above $E_F$, resulting in small electronic instabilities.[21, 22] On the other hand, almost all filled orbitals, as in $Cu_2OSeO_3$ ($d^9$), most *d*-states lie well below $E_F$, contributing little to electronic instabilities (Figure S11).

Exactly half-filled *d*- or *f*-orbitals tell us a unique story. These systems have large magnetic spin moments, with quenched orbital contributions; thus, they interact more directly with external magnetic fields via dipole interactions, and the influence of temperature is reduced. The external magnetic field couples directly to the spin moment, making the response to magnetic fields more pronounced and predictable. This behavior is evident in Eu- and Gd-based ($4f^7$) systems, as well as in MnSi and MnGe ($d^5$). When switching from Si ($3p$) to Ge ($4p$) in MnGe, the $E_{spin}$ remains largely unchanged, but the $E_{req}$ increases. This suggests that MnGe requires a greater energy input, through the magnetic field, than MnSi to transition to its skyrmion phase. This field sensitivity is expected from systems with predominantly spin-only magnetic moments (half-filled orbitals). In contrast, when the valence orbitals are slightly more than half-filled or spin-orbit coupling (SOC) is sizable, the total magnetic moment is a combination of spin and orbital angular momentum contributions ($J \neq S$). The resulting electronic states are defined by crystal-field splitting and closely spaced $J$-manifolds. These low-lying energy levels are sensitive to thermal excitations, and their population can change at finite temperature. In such systems, temperature plays a more important role in the state population.[92] For example, when Mn in MnGe ($d^5$) is replaced by Fe in FeGe ($d^6$),[93] the $E_{spin}$ and $E_{req}$ values of FeGe are higher than those of MnGe. This can explain why FeGe hosts skyrmions at higher temperatures than MnGe.

### 4.3. Influence of Electronic Structure in Metals and Insulators

Let us revisit **Figure 2**—the treasure map. Among the reported skyrmion materials, insulating systems have shown great promise for realizing the skyrmion phase at lower magnetic fields than metals. However, for utilities in spintronic applications, metallic materials offer increased tunability. In metals, skyrmions can be manipulated not only by magnetic fields but also by electric fields, enabling more refined control. Skyrmions in metallic systems may be stabilized by either DM or RKKY exchange. DM interactions arise in noncentrosymmetric lattices, while RKKY interactions require free charge carriers. Thus, metallic systems foster an additional driving force, mediated by conduction electrons, that can enhance or tune skyrmion formation. When targeting skyrmions at ambient conditions, the RKKY interaction alone is insufficient to provide the necessary driving force. In fact, some semi-metallic noncentrosymmetric magnets are able to stabilize DM-driven skyrmions close to room temperature at relatively low fields. Examples include FeGe (Bloch-type, stable up to 276 K at 150 Oe)[59], $Co_8Zn_9Mn_3$ (Bloch-type, 315 K at 1000 Oe)[60, 61], $Mn_2Rh_{0.95}Ir_{0.05}Sn$ (antiskyrmion, 237 K at 350 Oe)[62], and $Mn_{1.4}Pt_{0.9}Pd_{0.1}Sn$ (antiskyrmion, 400 K at 1650 Oe)[63]. These systems illustrate that skyrmions at ambient conditions are most likely to be realized in semi-metals with slightly more than half-filled $d$-orbitals ($d^6$, $d^7$). This observation suggests that looking into magnetic semi-metals that feature similar electronic instabilities and density of energy to those of the skyrmion hosts can be a jumping-off point toward the realization of skyrmions under ambient conditions.

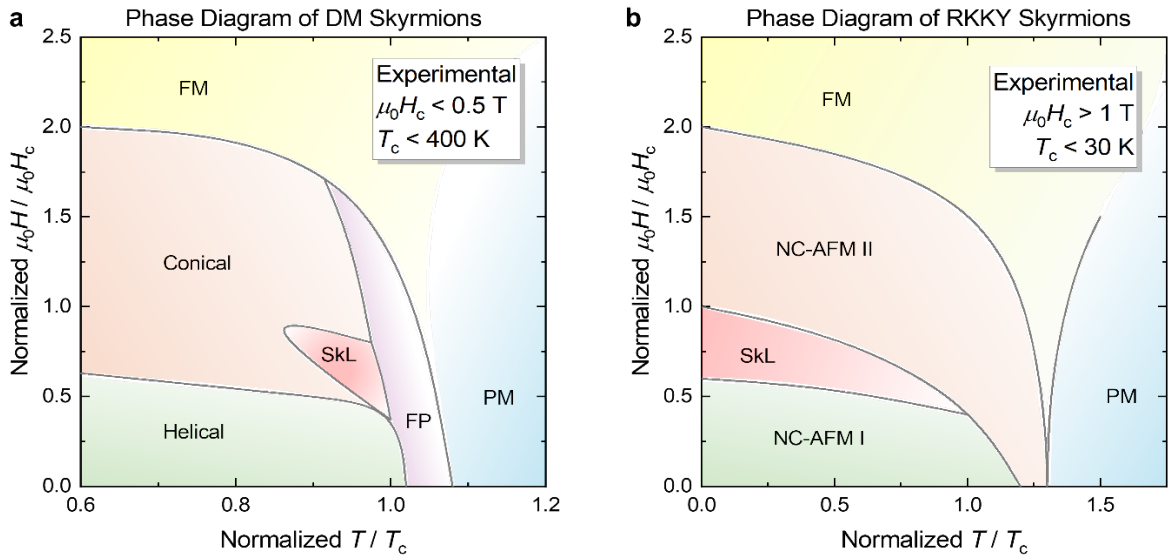

**Figure 9**. Normalized phase diagrams of DM- and RKKY-driven skyrmions. PM: paramagnetic; FP: field-polarized; FM: ferromagnetic; NC-AMF: non-collinear antiferromagnetic; SkL: skyrmions.

## 4.4. DM and RKKY Interactions

Within the aforementioned materials design perspectives, we have discussed the influence of *d*- vs f-electron systems, electron fillings in valence orbitals, and metals vs insulators. Now, let us return to the central Hamiltonians that govern the two underlying magnetic interactions responsible for skyrmion formation: the DM and the RKKY interaction. Understanding the nature of these Hamiltonians provides physical insight into why different skyrmion materials respond differently to external perturbations such as temperature and magnetic field. We constructed normalized magnetic phase diagrams corresponding to each interaction, enabling a direct visual comparison of the characteristic behavior of the two skyrmion classes (**Figure 9**).[20, 43, 48, 94-101]

The DM interaction originates from spin–orbit coupling in a non-centrosymmetric crystal structure. Its Hamiltonian is given by:

$$\mathcal{H}_{DMI} = \sum_{\langle i,j \rangle} \boldsymbol{D}_{ij} \cdot \left( \boldsymbol{S}_i \times \boldsymbol{S}_j \right) \qquad (3)$$

where $\boldsymbol{D}_{ij}$ is the DM vector, and $\boldsymbol{S}_i$, $\boldsymbol{S}_j$ are neighboring spins. This interaction favors non-collinear, chiral spin textures such as spin spirals and skyrmions. **Figure 9a** shows that as temperature decreases at zero field, a DM-driven skyrmion host goes through a paramagnetic state, then a field-polarized state and evolves into a helical ground state. While the DM interaction is necessary to stabilize these spin textures, its strength is typically small compared to the conventional Heisenberg exchange. In the presence of applied fields, the system also experiences the Zeeman effect. At modest fields, the Zeeman term induces spin alignment without disrupting the chiral winding, thereby stabilizing the conical spin configuration at low temperature or skyrmions at elevated temperature (**Figure 9a**). However, as the field increases beyond a critical threshold, the Zeeman energy overcomes the twisting of DM interaction, leading to a collapse of the skyrmion phase into a field-polarized state or a ferromagnetic state. It is worth noting that DM-driven skyrmions typically evolve at low fields across a broad range of temperatures up to 400 K (**Figure 2**, **Figure 9a**). Overall, the formation of DM-induced skyrmions can be described as a delicate balance among the Heisenberg exchange, the Zeeman effect, and the DM interaction, consistent with phase diagrams in which the skyrmion phase occupies only a narrow region between other conventional magnetic states.

In contrast, the RKKY interaction, which arises from indirect exchange coupling mediated by conduction electrons, is strongly dependent on the electronic structure. Its simplified form is:

$$\mathcal{H}_{RKKY} \propto J_{RKKY}(r)\boldsymbol{S}_i \cdot \boldsymbol{S}_j \qquad (4)$$

$$J_{RKKY}(r) \sim \frac{\cos(2k_F r)}{r^3} \qquad (5)$$

The coupling strength $J_{RKKY}$ is an oscillatory function of the interatomic distance $r$ and the Fermi wavevector $k_F$. RKKY exchange—a first-order effect in exchange coupling—can exceed the magnitude of the DM interaction (a second-order effect). This difference in interaction strength is reflected in the distinct ground states of skyrmion materials (**Figure 9**). DM-driven hosts typically exhibit a helical ground state, in which neighboring spins rotate by less than 180° due to the influence of the DM vector. In contrast, some RKKY-driven hosts display AFM or incommensurate non-collinear AFM ground states, indicating that $J_{RKKY}$ can be sufficiently strong to reverse a spin from its neighbor.

RKKY-driven skyrmion evolution is particularly susceptible to the topology of the Fermi surface. A vital feature that facilitates skyrmion formation in magnetic metals is Fermi surface nesting—large, parallel sections of the Fermi surface that lead to anomalies in the electronic susceptibility at certain wavevectors. However, this nesting is a low-temperature phenomenon; as temperature increases, the Fermi–Dirac distribution becomes broader, smearing out these features and reducing the effectiveness of the spin-mediated coupling. Thus, elevated temperatures can suppress the underlying mechanism that stabilizes RKKY-induced skyrmions. In fact, RKKY-driven skyrmions typically emerge at high fields and low temperatures (**Figure 2**). But once the skyrmions are stabilized, their presence extends from $T_c$ down to base temperature (**Figure 9b**).

Taken together, the discussion on the underlying DM and RKKY interactions illuminates, in part, some insight into why DM-driven skyrmions tend to evolve over a wide temperature window at low fields, while RKKY-induced skyrmions typically form at low temperatures and high fields (**Figure 2**, **Figure 9**). This framework aligns well with the observations discussed in Sections 4.1 through 4.3. For instance, the different field and temperature conditions seen in $d$- and $f$-electron systems can now be better understood by considering both DOE and magnetic exchange interactions. This understanding could further guide design strategies for skyrmion hosts at ambient conditions by providing some hints of the electronic origins that foster a phase transition to skyrmions.

### 4.5. Implications for Skyrmion Discovery and Design Strategies

The perspective highlights the importance of tailoring the electronic structure, through careful magnetic element selection, strategic electron fillings of valence orbitals, and desired bonding character, as a guiding strategy for choosing skyrmion materials of interest. This framework offers a predictive pathway for evaluating and engineering systems that stabilize skyrmions under accessible field and temperature conditions.

One important factor not addressed in depth here is the size of skyrmions, which critically impacts their stability, dynamics, and suitability for device uses. It has been experimentally observed that DM-driven skyrmions often have larger sizes (a few to hundreds of nm) compared to those stabilized by RKKY-type interactions (a few nm). Moreover, another salient experimental observation is that skyrmion sizes are temperature-dependent. RKKY-driven skyrmions, typically evolving at higher magnetic fields ($\mu_0 H_c > 1$ T) and lower temperatures, display small sizes. DM-induced skyrmions emerging at higher temperatures are typically larger in size than those forming at low temperatures, suggesting that thermal excitations are at play. Although many excellent studies have modeled skyrmion sizes using micromagnetic simulations or continuum models, the microscopic origin of skyrmion sizes still remains to be further investigated.[102-107]

In this perspective, DOE and $DOE_s$ analysis is based on spin-polarized DFT that studies the ground-state electronic structure. While this approach provides valuable insights into electronic instabilities that can give rise to a skyrmion phase transition, it cannot resolve the spatial size or profile of skyrmions. As mentioned earlier, noncollinear DFT calculations using supercells that explicitly incorporate the skyrmion spin configuration would allow for direct comparison of total energies, electronic structure, and spin texture features, offering more quantitative guidance to link the electronic structure of the excited state and skyrmion sizes.

Nevertheless, this approach provides an effective framework for judiciously tuning the electronic structure of the material to promote susceptibility to skyrmion evolution. Let us take $Mn_2RhSn$ as an example. This pristine compound does not host skyrmions; however, its electron-doped systems $Mn_2Rh_{0.95}Ir_{0.05}Sn$ and $Mn_{1.4}Pt_{0.9}Pd_{0.1}Sn$ stabilize skyrmions at high temperatures. This occurs because electron doping positions the Fermi level into a region where electronic instabilities are amplified.[108, 109] Another example that also benefits from the approach is the B20 compounds (e.g., FeSi, MnGe). Hole-doping these systems with Co or Cr, such as $Fe_{1-x}Co_xSi$ and $Cr_{0.82}Mn_{0.18}Ge$, results in the evolution of skyrmions at a wide temperature range and low fields. The electronic tunability approach resonates with other excellent studies in the field.[49, 110-114]

## 5. Summary and Outlook

In this perspective, we have articulated a unified framework of the electronic origins underpinning skyrmion formation across a diverse set of materials. The material examples in this study are established skyrmion hosts; thus, they already possess a pathway that enables a magnetic phase transition to skyrmions. By examining $E_{spin}$ and $E_{req}$ derived from the DOE and $DOE_s$ and experimental data, a semi-quantitative model for linking the electronic instability and the temperature and field conditions of skyrmion evolution is developed. We demonstrate that this approach can be applied across RKKY- and DM-driven systems, spanning metallic and insulating hosts with $d$- and $f$-electron magnetism. Normalizing $E_{spin}$ and $E_{req}$ by a unit-cell volume enables a meaningful comparison for the diverse set of skyrmion materials. We then discuss broader design considerations, highlighting how orbital character, electron fillings, magnetic exchange, and structural factors influence a skyrmion phase transition. In addition, the approach can serve as a predictive tool: (i) doping can turn a non-skyrmion host into a skyrmion host by positioning the Fermi level into a regime in the DOE where electronic instabilities are enhanced (modifying $E_{spin}$ and $E_{req}$), and (ii) for known skyrmion hosts, the $DOE_s$ curve can also be used as a guide to tune skyrmion formation conditions by modifying the electronic instability through chemical adjustment.

The normalized phase diagrams highlight distinct characteristics of skyrmion regions stabilized by DM or RKKY interactions. DM-driven skyrmions are confined to a narrow region between conventional magnetic states, whereas RKKY-induced skyrmions typically span a broad range, extending from the critical temperature down to the base temperature. Moreover, DM-driven skyrmions can evolve over a wide temperature window at low fields, while RKKY-driven skyrmions generally emerge only at high fields and low temperatures.

Taking all into consideration, we predict that noncentrosymmetric semi-metals with slightly above half-filled $d$-orbitals would be excellent candidates for hosting skyrmions at ambient conditions (the treasure in **Figure 2**). The framework elucidates the electronic origins governing skyrmion evolution and establishes a useful link between these insights and experimentally measurable parameters—specifically, the temperature and field conditions of a skyrmion phase transition. This perspective shares a new way of approaching quantum materials research that could assist researchers in advancing the design and development of skyrmions at ambient conditions, while appreciating the current remarkable research efforts in the field that have built the foundation necessary for many more forthcoming, tantalizing breakthroughs now within reach.

ASSOCIATED CONTENT

**Supporting Information**.

AUTHOR INFORMATION

**Corresponding Author**

Email: thao@clemson.edu

**Author Contributions**

The manuscript was written through the contributions of all authors. All authors have given approval to the final version of the manuscript.

**Funding Sources**

This work was supported by the Arnold and Mabel Beckman Foundation grant 2023 BYI and the Camille and Henry Dreyfus Foundation grant 2025 Camille Dreyfus Teacher-Scholar award TC-25-071. The synthesis of some of the skyrmions was supported by the U.S. National Science Foundation grant NSF-DMR-CAREER-2338014.

**Notes**

Any additional relevant notes should be placed here.

ACKNOWLEDGMENT

We thank Athira Babu and Abigail Corales for their assistance in the synthesis of two skyrmion host materials.

Supplementary Materials for

**Treasure Map Toward Skyrmion Evolution in Ambient Conditions:**

**A Perspective from Electronic Instabilities and the Density of Energy**

Xudong Huai and Thao.T. Tran*

Corresponding author: thao@clemson.edu

**The PDF file includes:**

**Experimental Section**

## Experimental Section

### Synthesis

Polycrystalline VOSe$_2$O$_5$ was prepared via a conventional solid state reaction. VO$_2$ and SeO$_2$ were weighed in stoichiometric amounts, thoroughly ground using an agate mortar and pestle, and pressed into pellets. The pellets were sealed under vacuum in quartz tube and annealed at 400 °C for 96 h, with 3 intermediate regrinding steps to enhance homogeneity. The resulting pale-green powder was confirmed single-phase by powder X-ray diffraction

For Cu$_2$OSeO$_3$, stoichiometric CuO and a slight molar excess of SeO$_2$ were similarly mixed, ground, and pelletized. The pellets were vacuum-sealed in quartz tube and heated at 600 °C for 72 h, including 3 intermittent grinding stage to ensure complete reaction. Phase-pure olive-green powders were verified by powder X-ray diffraction.

### Magnetization

DC magnetization measurements on VOSe$_2$O$_5$ and Cu$_2$OSeO$_3$ powder samples were conducted using the Vibrating Sample Magnetometer option of the Quantum Design Magnetic Properties Measurement System (MPMS-3). The magnetic entropy change ($\Delta S_{mag}$) was subsequently extracted from temperature-dependent magnetization data by evaluating d$M$/d$T$ across field sweeps and integrating using the Maxwell relation.

### Density Functional Theory Calculation

Pseudo-potential DFT calculations were conducted using Quantum Espresso (QE) software package[1] in conjunction with the Generalized Gradient Approximation (GGA+U)[2] for the exchange-correlation potential, specifically adopting the PBE parametrization[3]. For the projector-augmented wave (PAW) potentials, all elements were sourced from the PSlibrary v.1.0.0 dataset,[4] the k-mesh employed and kinetic energy cutoffs for both charge density and wavefunctions were optimized for each Skyrmion hosts, and the computation details are summarized in Table S1. To elucidate electronic structure, the spin-resolved density of energy were computed using the LOBSTER software.[5, 6]

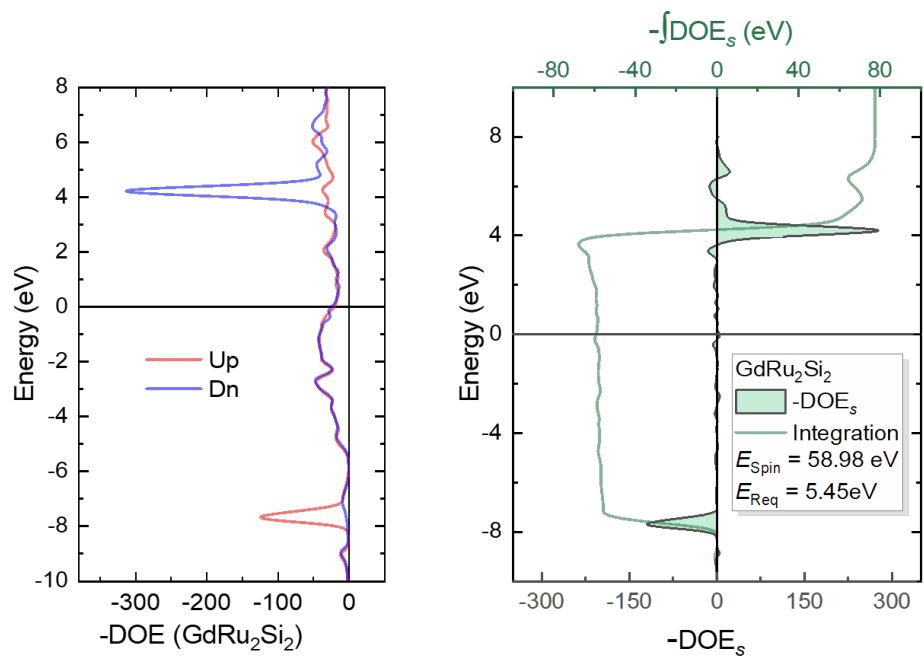

**Figure S1.** Density of energy for GdRu$_2$Si$_2$: (a) spin-resolved density of energy (DOE); (b) spin asymmetry (DOE$_s$) and its integration.

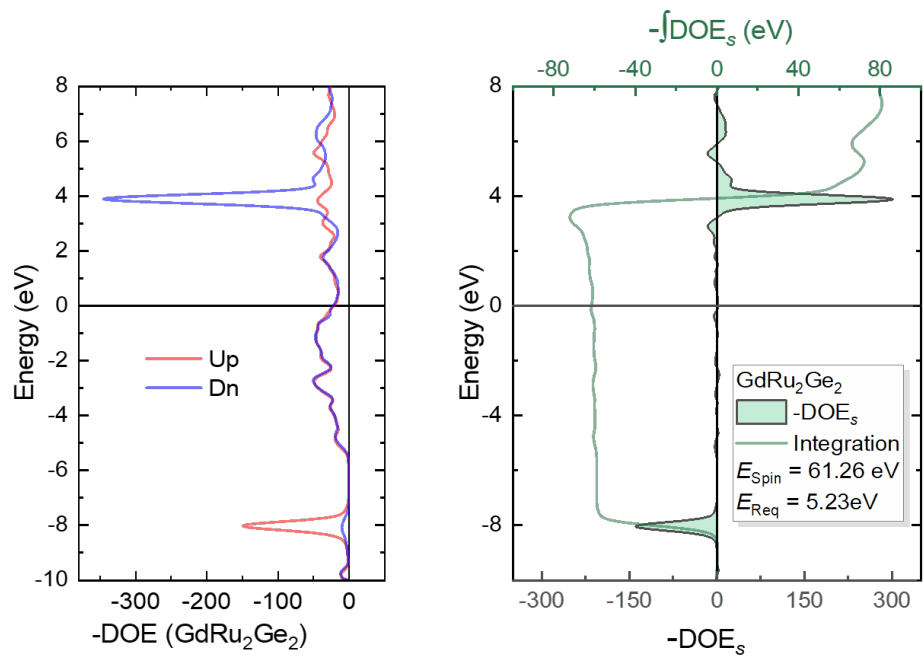

**Figure S2.** Density of energy for GdRu$_2$Ge$_2$: (a) spin-resolved density of energy (DOE); (b) spin asymmetry (DOE$_s$) and its integration.

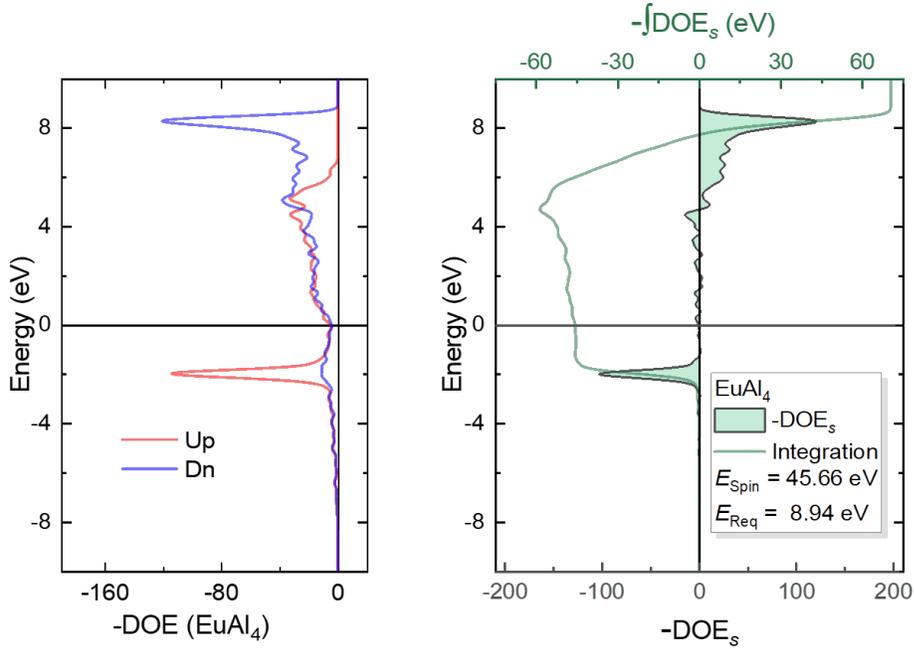

**Figure S3.** Density of energy for EuAl$_4$: (a) spin-resolved density of energy (DOE); (b) spin asymmetry (DOE$_s$) and its integration.

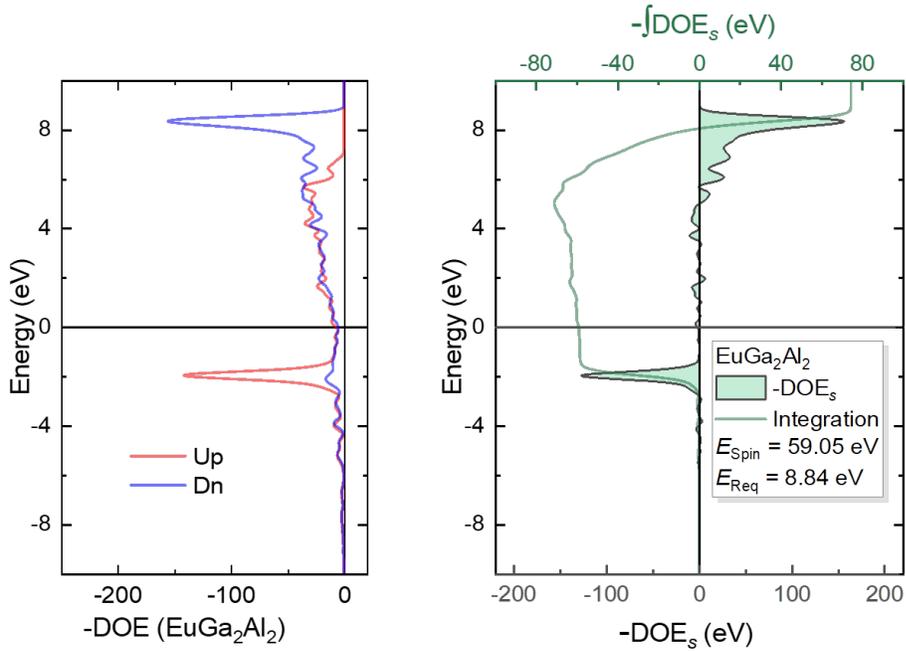

**Figure S4.** Density of energy for EuGa$_2$Al$_2$: (a) spin-resolved density of energy (DOE); (b) spin asymmetry (DOE$_s$) and its integration.

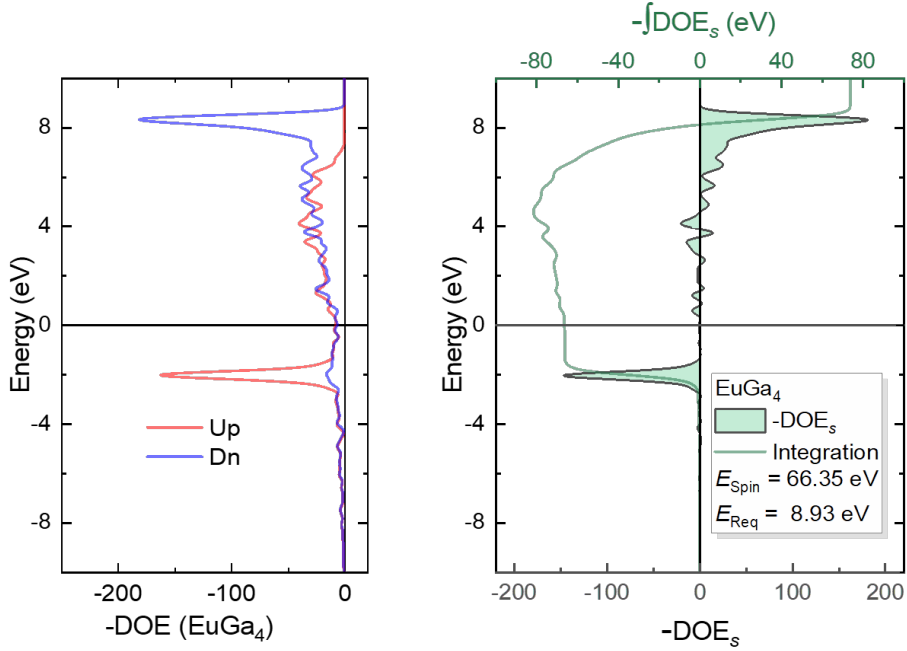

**Figure S5.** Density of energy for EuGa$_4$: (a) spin-resolved density of energy (DOE); (b) spin asymmetry (DOE$_s$) and its integration.

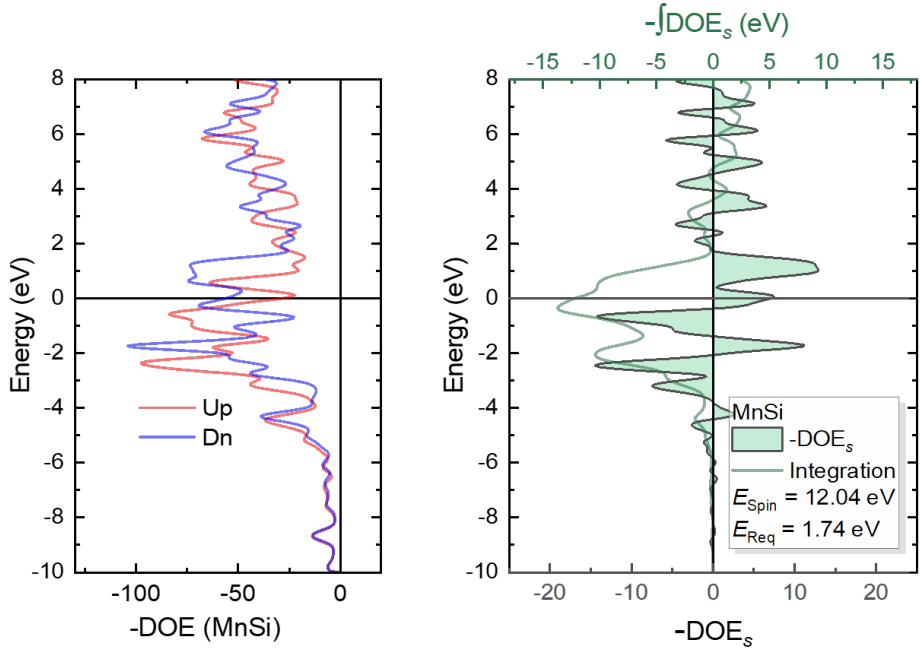

**Figure S6.** Density of energy for MnSi (a) spin-resolved density of energy (DOE); (b) spin asymmetry (DOE$_s$) and its integration.

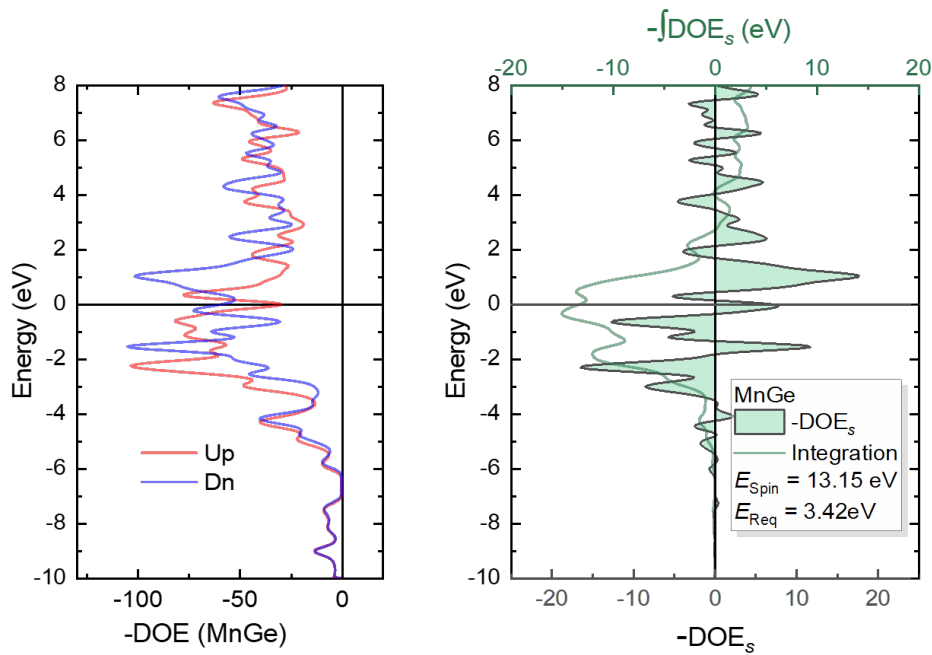

**Figure S7.** Density of energy for MnGe: (a) spin-resolved density of energy (DOE); (b) spin asymmetry (DOE$_s$) and its integration.

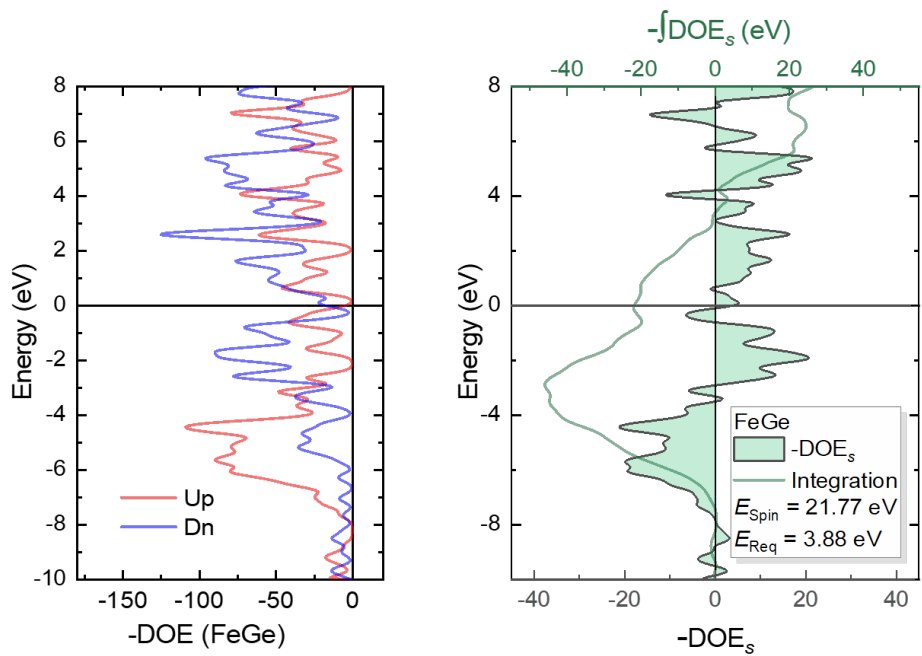

**Figure S8.** Density of energy for FeGe: (a) spin-resolved density of energy (DOE); (b) spin asymmetry (DOE$_s$) and its integration.

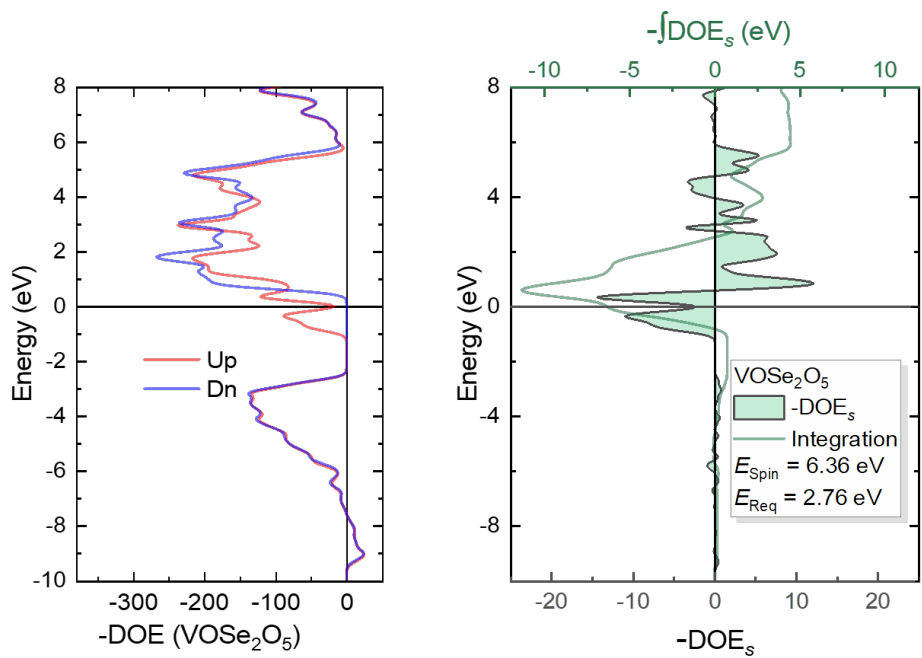

**Figure S9.** Density of energy for VOSe$_2$O$_5$: (a) spin-resolved density of energy (DOE); (b) spin asymmetry (DOE$_s$) and its integration.

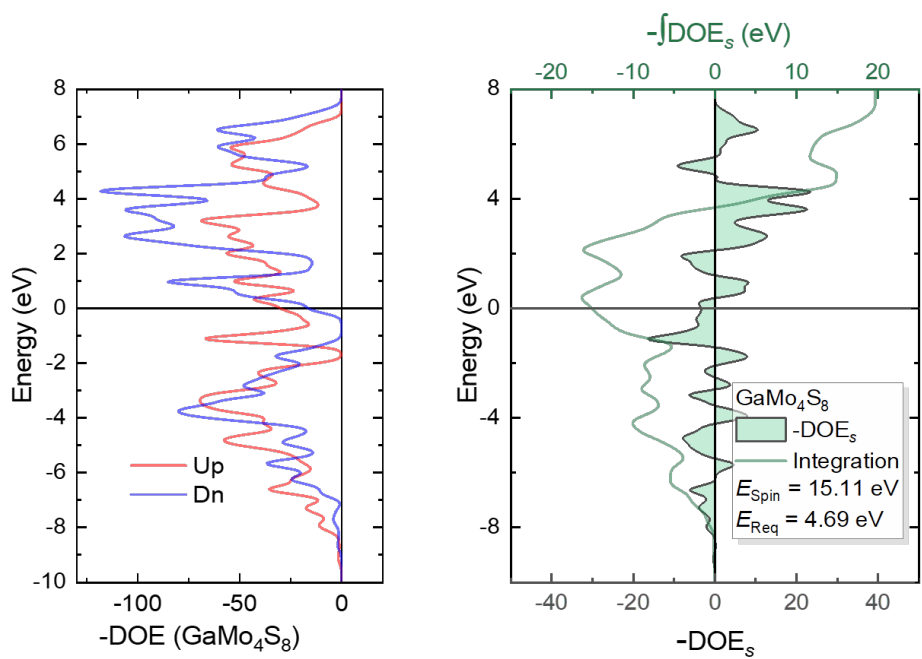

**Figure S10.** Density of energy for GaMo$_4$S$_8$: (a) spin-resolved density of energy (DOE); (b) spin asymmetry (DOE$_s$) and its integration.

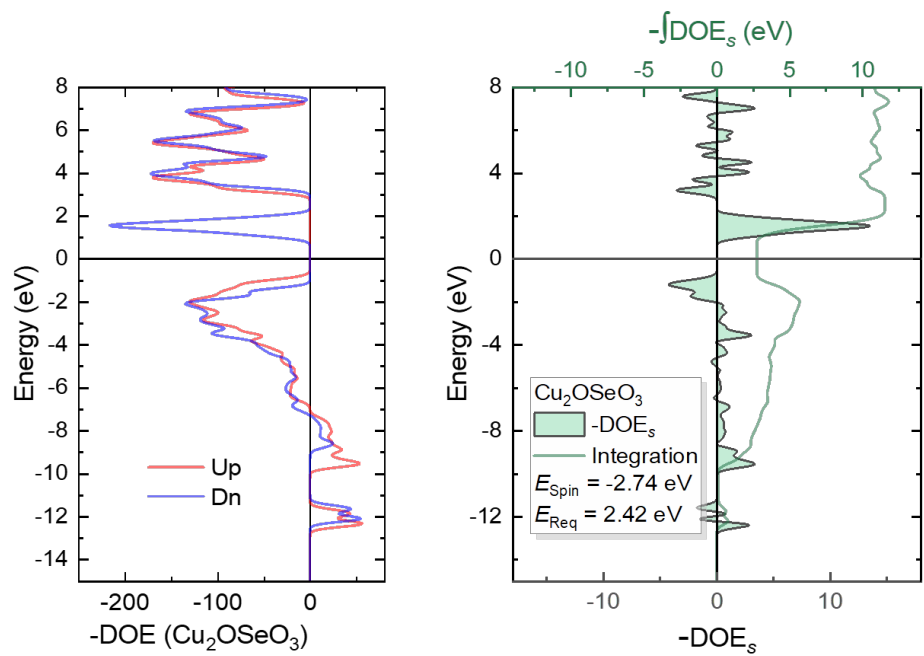

**Figure S11.** Density of energy for $Cu_2OSeO_3$: (a) spin-resolved density of energy (DOE); (b) spin asymmetry ($DOE_s$) and its integration.

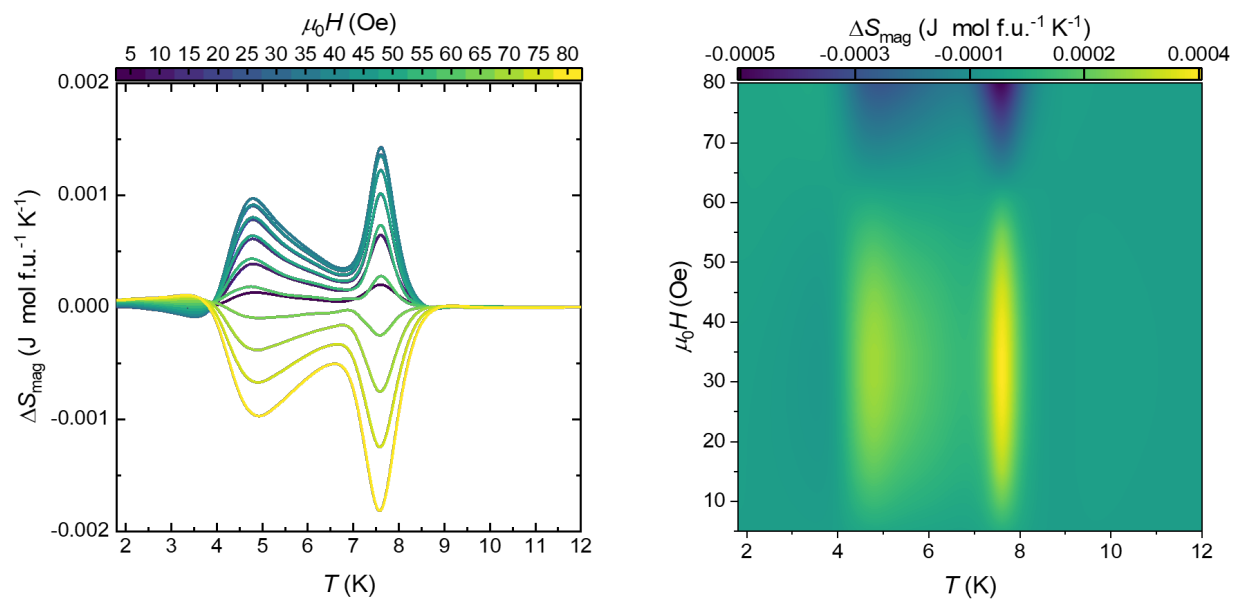

**Figure S12.** Magnetic entropy of $VOSe_2O_5$.

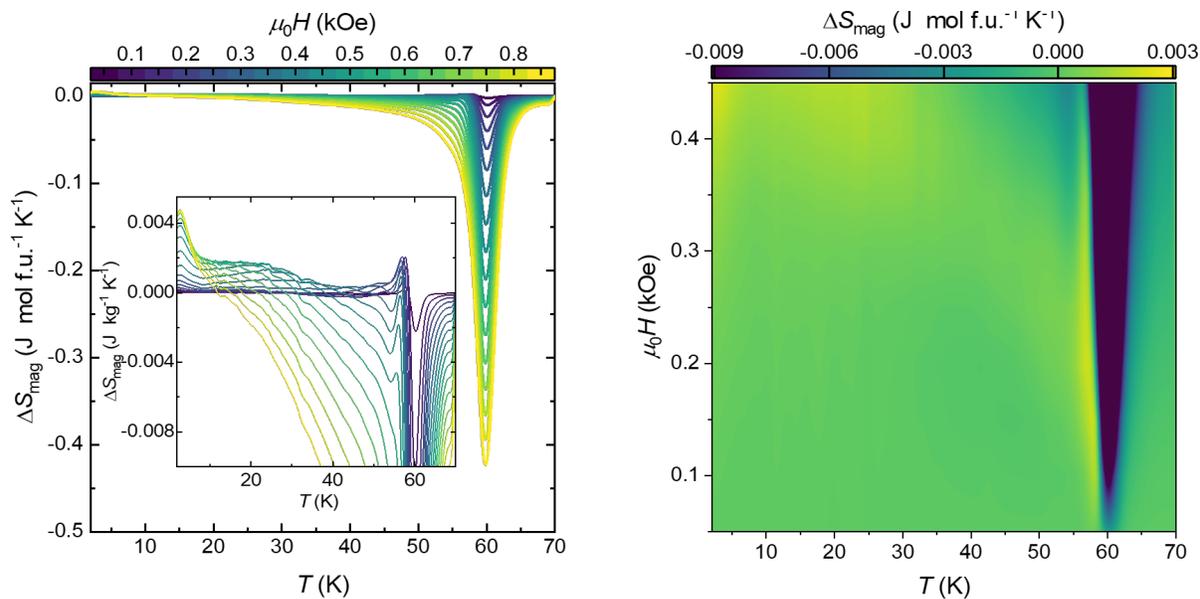

**Figure S13.** Magnetic entropy of $Cu_2OSeO_3$.

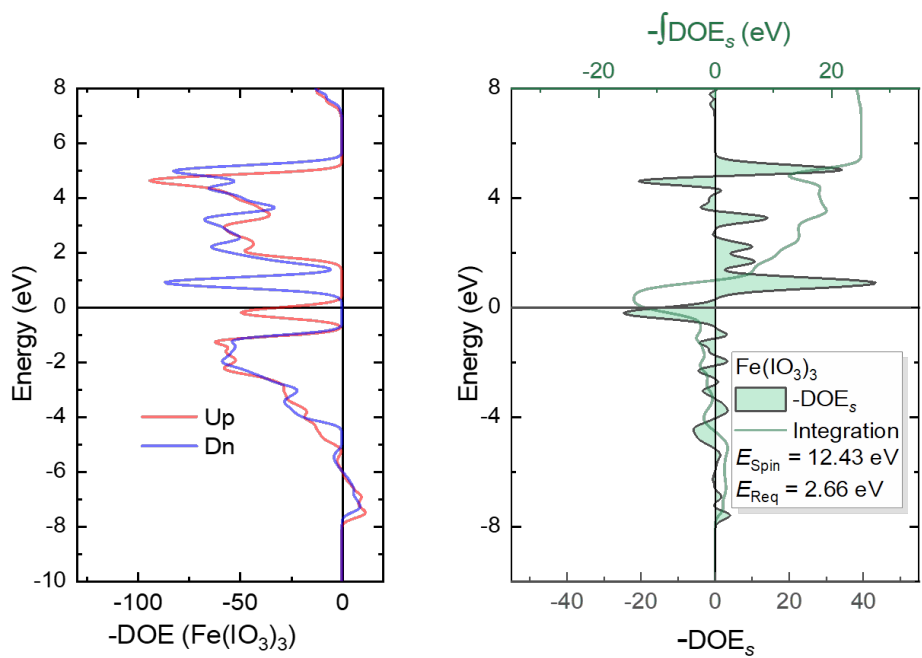

**Figure S14.** Density of energy for $Fe(IO_3)_3$: (a) spin-resolved density of energy (DOE); (b) spin asymmetry ($DOE_s$) and its integration.

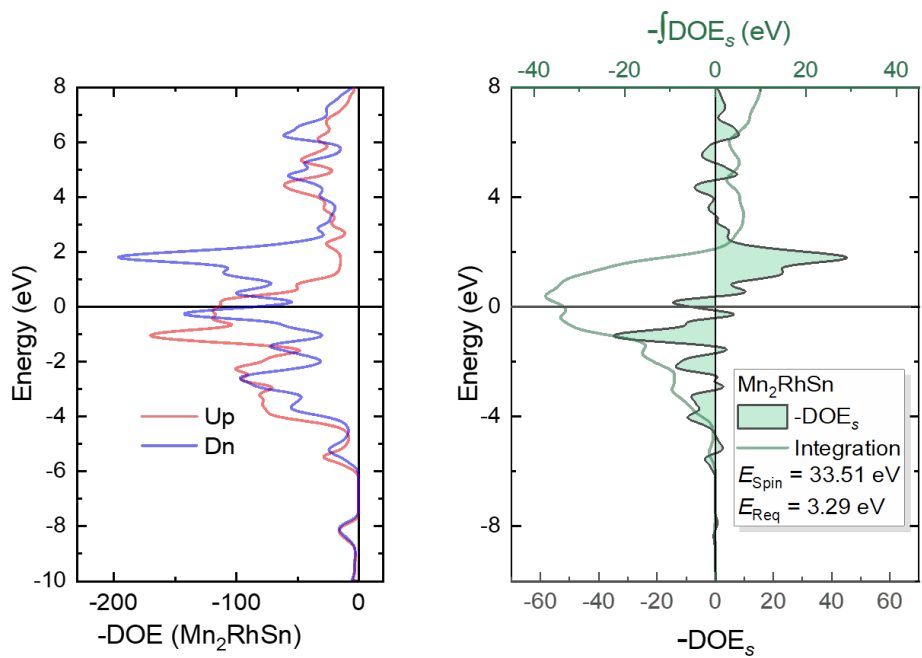

**Figure S15.** Density of energy for Mn₂RhSn: (a) spin-resolved density of energy (DOE); (b) spin asymmetry (DOE$_s$) and its integration.

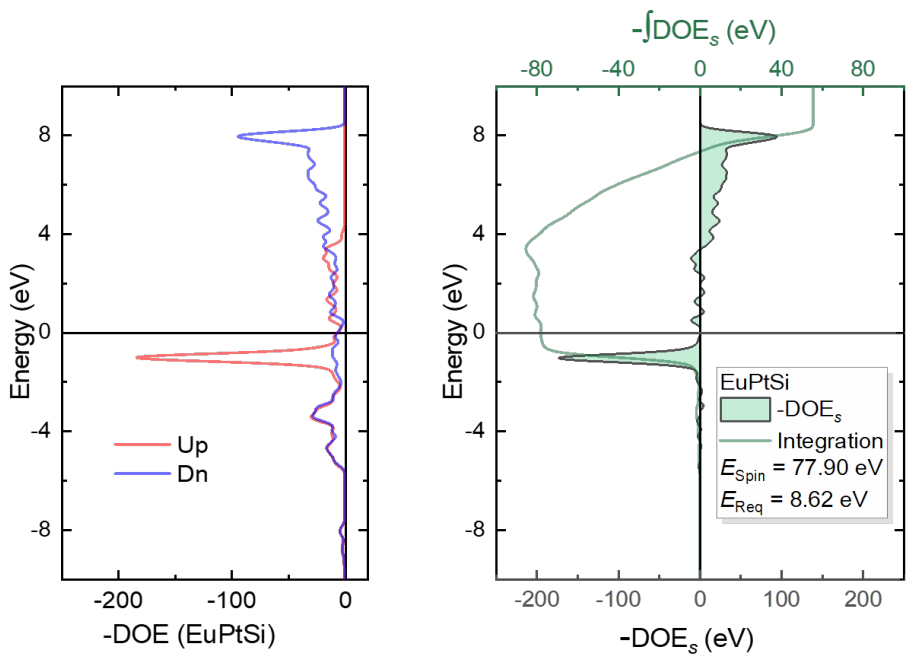

**Figure S16.** Density of energy for EuPtSi: (a) spin-resolved density of energy (DOE); (b) spin asymmetry (DOE$_s$) and its integration.

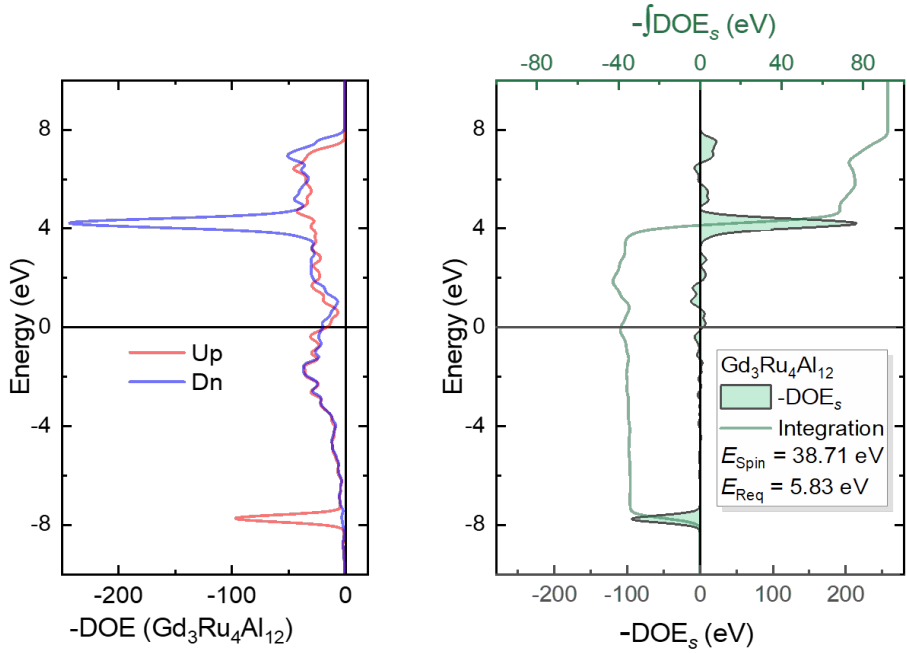

**Figure S17.** Density of energy for Gd$_3$Ru$_4$Al$_{12}$: (a) spin-resolved density of energy (DOE); (b) spin asymmetry (DOE$_s$) and its integration.

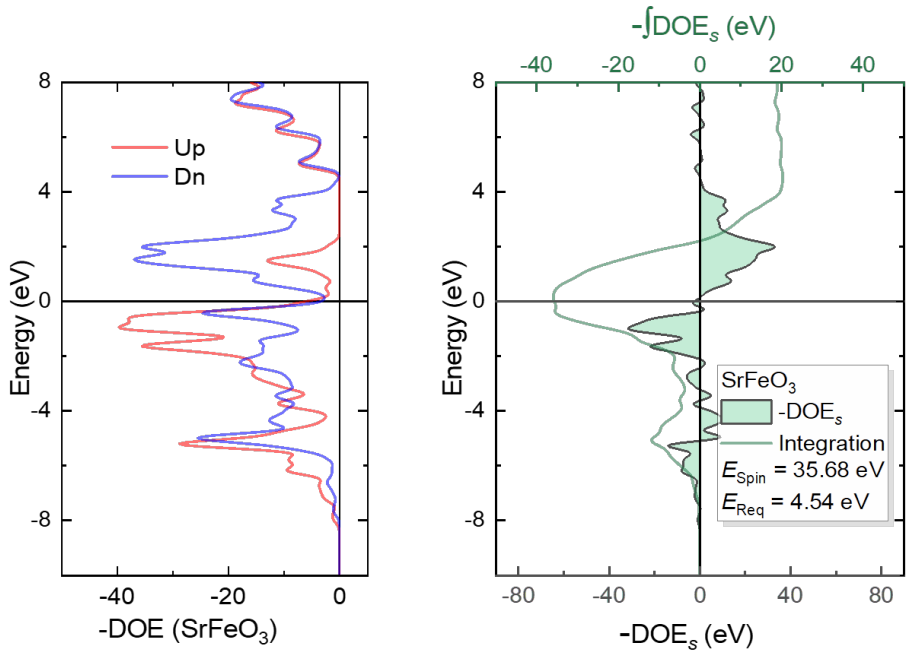

**Figure S18.** Density of energy for SrFeO$_3$: (a) spin-resolved density of energy (DOE); (b) spin asymmetry (DOE$_s$) and its integration.

**Semi-Quantitative Assessment and Experimental Correlation**

**Pure theoretical approach**

If the skyrmion formation is actually closely related to the available destabilized excited states in the electronic structure, we would expect the energy provided by the field and temperature to be close to the energy required for the skyrmion formation. In order to cover more skyrmion candidates, we used a pure theoretical model of the external energy ($E_{ex}$) that can be extended to most of the hosts:

$$E_{ex}^{M1} = \frac{9\,Nk_BT^4}{\theta_D{}^3} + \frac{H^2V}{\mu_0\mu} \qquad (S1)$$

The first term describes the thermal energy input arising from phonon excitations. Here, $N$ is the number of atoms per $\text{Å}^3$. One may intend to use the number of magnetic atoms, but the $E_{req}$ is extracted from spin-asymmetry DOE$_s$; this characteristic value still represents the material instability change. $k_B$ is Boltzmann's constant, $T$ is the absolute temperature, and $\theta_D$ is the Debye temperature, which sets the energy scale of lattice vibrations. The $\theta_D$ were fitted and fixed at 135 K for all Skyrmion hosts, which is far from accurate but sufficient for a rough validation. Within the low-temperature limit ($T \ll \theta_D$), this $T^4$-dependence stems from the Debye model and quantifies the contribution of thermal fluctuations to destabilizing the ground-state electronic and magnetic order. The second term accounts for the magnetic energy supplied by an external field $H$, where $V$ is the unit-cell volume, $\mu_0$ is the vacuum permeability, and $\mu$ is the magnetic permeability of the material. This term corresponds to the magnetic work that can overcome exchange interactions and drive the system into an excited magnetic state, such as a skyrmion phase. A fitted $\mu_{\text{fitted}} = 0.26$ H/m was used for all hosts, again, far from accurate, just to approach the magnitude. Together, these two contributions describe the total energy available to promote the formation of skyrmions under finite temperature and magnetic field.

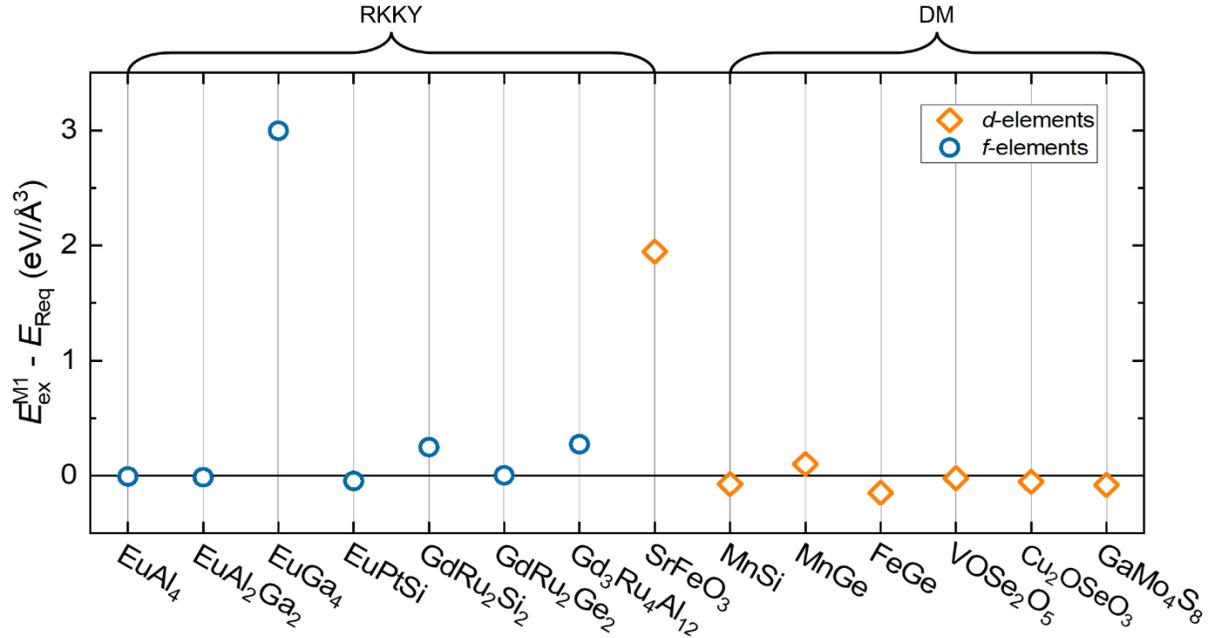

**Figure S19**. Difference between the calculated external energy ($E_{ex}$) using model 1 and the extracted excitation energy ($E_{req}$) for 14 skyrmion hosts.

The calculated external energy was then compared to the extracted $E_{req}$. (Figure S19) Given that $E_{req}$ lies in the order of $10^{-2}$ to $10^{-1}$ eV/Å$^3$, this model successfully captures the required energy scale for several systems. However, it clearly overestimates the magnetic-field contribution, as hosts requiring $H_{Smax} > 1$ T (e.g. EuGa$_4$, SrFeO$_3$, and MnGe)[7-9] exhibit energy scales far exceeding $E_{req}$. (Figure 7b) In addition, the thermal contribution is underestimated for hosts with $T_{Smax} > 11$ K, such as MnSi, FeGe, and GaMo$_4$S$_8$. These discrepancies may be improved with more accurate values of $\theta_D$ and $\mu$, but the agreement between calculated $E_{ex}$ and $E_{req}$ for low-field and low-temperature systems supports the idea that further destabilization of states above $E_F$ is closely linked to skyrmion formation.

**Experimental data supported approach**

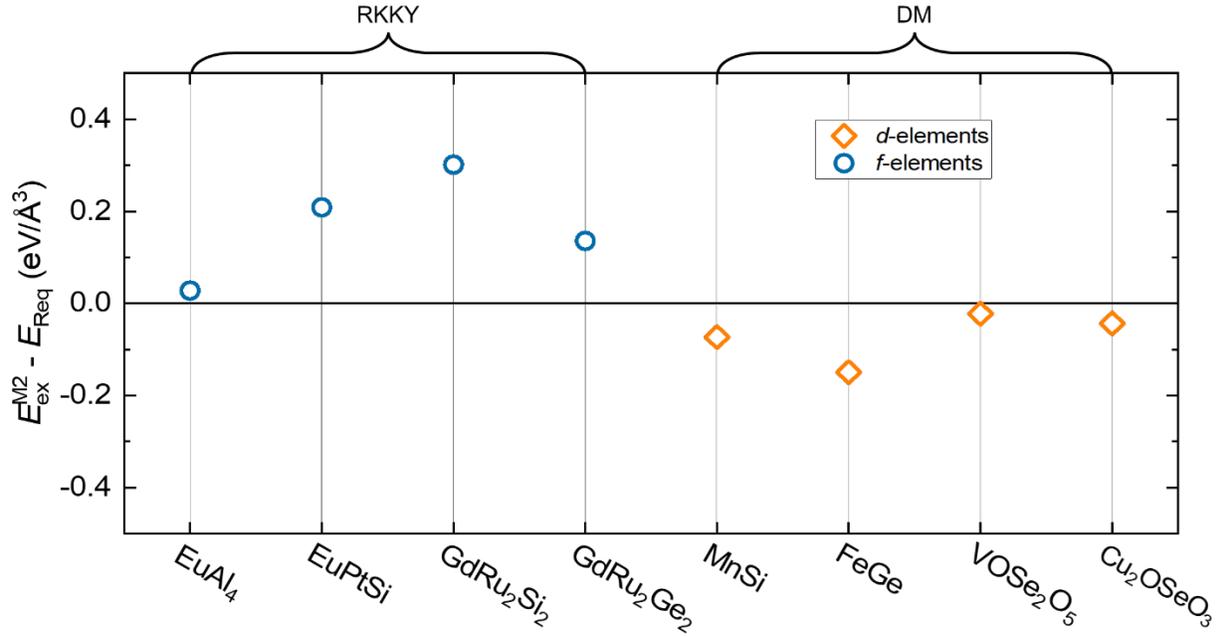

**Figure S20**. Difference between the calculated external energy ($E_{ex}$) using model 2 and the extracted excitation energy ($E_{req}$) for 8 skyrmion hosts.

Since the theoretical approach indicates the need for accurate data, we then propose the second model which is data-based for the external energy supplied to the system:

$$E_{ex}^{M2} = \frac{T \, \Delta S_{mag} \, m_{cell}}{M V_{cell}} + \frac{1}{2} M H V \qquad (S2)$$

The first term represents the thermal energy contribution derived from the magneto-entropy $\Delta S_{mag}$, scaled by temperature $T$, magnetic moment per unit cell $m_{cell}$, molar mass $M$, and unit-cell volume $V_{cell}$. This quantifies the thermal energy needed to reach the magnetically excited state observed in magneto-entropy analysis from magnetization of heat capacity measurements. The second term corresponds to the magnetic energy supplied by the external field $H$, where $M$ is the magnetization per unit volume. Unlike the previous, more theoretical model, this formulation is grounded in direct experimental parameters (e.g. $\Delta S_{mag}$, $M$, $V$, $H$) and thus provides a practical means to estimate the energy scale driving skyrmion formation.

The calculated external energy was also compared with $E_{req}$ (Figure S20), and this second model reproduces the required energy scale ($10^{-2}$ to $10^{-1}$ eV/Å$^3$) much more accurately. Although the limited availability of data restricts the number of compounds considered, the model successfully captures the behavior of EuAl$_4$, MnSi, and FeGe. This further supports the idea that the destabilization of states above $E_F$ plays a central role in skyrmion formation.

**Proposed approach**

Despite the improvements in the second model, we still observe a ~30% or greater difference between $E_{req}$/Å$^3$ and $E_{ex}$/Å$^3$. This discrepancy may be partly due to the assumptions made in the ground-state calculation. Specifically, we relied on collinear spin-polarized DFT and performed a spin-resolved DOE analysis under this constraint. However, several of the compounds considered do not exhibit a classical collinear magnetic ground state. For instance, EuAl$_4$ adopts a collinear AFM order, but its magnetic structure is incommensurate with the lattice,[10] and other examples such as MnSi, FeGe, EuPtSi, and Cu$_2$OSeO$_3$ are known to host helical magnetic ground states.[11-15] These departures from conventional FM or AFM ordering may themselves originate from the electronic instability near $E_F$. Furthermore, in the excited states under finite temperature or field, the electronic structure may be substantially altered relative to the ground-state configuration. Hence, extracting $E_{req}$ solely from the ground-state band structure above $E_F$ is likely to be an oversimplification.

A more rigorous approach would involve noncollinear DFT calculations with supercells that reflect both the true magnetic structure of the ground-state and the skyrmion (excited-state) configuration. The normalized total energy difference between these two states would then provide a more accurate estimate of $E_{\text{req}}$. However, such calculations are computationally expensive, especially if we aim to apply this methodology to large materials databases in search of new skyrmion candidates. Nevertheless, the present model already captures the key trends and provides a reasonable starting point for rapid screening.

**Table S1**: DFT details: k-mesh, Hubbard U, and Energy cutoff

| Skyrmion Hosts | k-mesh (a×b×c) | Charge Density Cutoff (Ry) | Wavefunction Cutoff (Ry) | Hubbard Parameter (eV) |
|---|---|---|---|---|
| $Cu_2OSeO_3$ | 4×4×4 | 497 | 55 | Cu-3$d$ 9.0 eV |
| FeGe | 3×3×3 | 411 | 45 | N/A |
| $GaMo_4S_8$ | 4×4×4 | 306 | 39 | Mo-4$d$ 5.0 eV |
| MnGe | 5×5×5 | 225 | 25 | Mn-3$d$ 5.0 eV |
| MnSi | 5×5×5 | 250 | 40 | Mn-3$d$ 5.0 eV |
| $VOSe_2O_5$ | 2×2×3 | 641 | 47 | N/A |
| $EuAl_4$ | 6×6×3 | 1099 | 122 | Eu-4$f$ 5.7 eV |
| $EuGa_4$ | 6×6×3 | 1099 | 122 | Eu-4$f$ 5.7 eV |
| $EuGa_2Al_2$ | 6×6×3 | 1099 | 122 | Eu-4$f$ 5.7 eV |
| EuPtSi | 4×4×4 | 1099 | 122 | Eu-4$f$ 4.5 eV |
| $Gd_3Ru_4Al_{12}$ | 6×6×4 | 1049 | 116 | Gd-4$f$ 6.7 eV |
| $GdRu_2Ge_2$ | 6×6×4 | 734 | 45 | Gd-4$f$ 6.7 eV |
| $GdRu_2Si_2$ | 6×6×4 | 734 | 45 | Gd-4$f$ 6.7 eV |
| $SrFeO_3$ | 6×6×6 | 225 | 25 | N/A |

**Table S2**: Skyrmion hosts, formation conditions and calculated energy required and spin energy.

| Skyrmion Hosts | $T_{max}$ (K) | $H_{max}$ (T) | Volume (Å³) | $E_{Req}$ (eV) | $E_{Req}$ per Å³ | $E_{spin}$ per Å³ | Size (nm) | Ref |
|---|---|---|---|---|---|---|---|---|
| $Cu_2OSeO_3$ | 2.7 | 0.075 | 710.93 | 2.42 | 0.054 | -0.99 | 62 | 16-18 |
| FeGe | 276 | 0.025 | 103.82 | 3.88 | 0.149 | 3.35 | 70 | 19 |
| $GaMo_4S_8$ | 25 | 0.1 | 230.08 | 4.69 | 0.082 | 1.05 | | 20-22 |
| MnGe | 20 | 2.4 | 110.07 | 3.42 | 0.124 | 1.91 | 3-6 | 23 |
| MnSi | 28 | 0.12 | 95.21 | 1.74 | 0.073 | 2.02 | 18 | 12, 24 |
| $VOSe_2O_5$ | 7.6 | 0.003 | 989.48 | 2.76 | 0.022 | 0.41 | 140 | 25, 26 |
| $EuAl_4$ | 10 | 1 | 215.57 | 8.94 | 0.083 | 0.85 | | 27, 28 |
| $EuGa_4$ | 2 | 6.5 | 205.71 | 8.93 | 0.087 | 1.29 | | 29 |
| $EuGa_2Al_2$ | 2 | 1 | 204.02 | 8.84 | 0.087 | 1.16 | | 30 |
| EuPtSi | 4.3 | 1 | 266.59 | 9.4 | 0.141 | 4.68 | 1.8 | 15, 31 |
| $Gd_3Ru_4Al_{12}$ | 5 | 1.2 | 641.30 | 5.83 | 0.055 | 2.17 | 2.8 | 32 |
| $GdRu_2Ge_2$ | 20 | 1 | 176.67 | 5.23 | 0.059 | 1.39 | | 33 |
| $GdRu_2Si_2$ | 10 | 2.3 | 166.73 | 5.47 | 0.066 | 1.41 | | 34, 35 |
| $SrFeO_3$ | 105 | 10 | 57.11 | 4.54 | 0.079 | 0.62 | 1.8 | 7-9 |